\documentclass{aa}
\usepackage{newtxtext,newtxmath}
\usepackage{booktabs}
\usepackage[T1]{fontenc}

\DeclareRobustCommand{\VAN}[3]{#2}
\let\VANthebibliography\thebibliography
\def\thebibliography{\DeclareRobustCommand{\VAN}[3]{##3}\VANthebibliography}

\usepackage{graphicx}	% Including figure files
\usepackage{amsmath}	% Advanced maths commands

\begin{document}

   \title{On the detection of Population III galaxies: Emission Line Diagnostics for Hybrid Stellar Populations}

    \author{S. Goswami
          \inst{1}
          \and
          J. M. Vílchez\inst{1}
          \and
          C. Kehrig\inst{1,2}
          \and
          A. Ferrara\inst{3}
          \and
          J. Iglesias-Páramo\inst{1,4}
          }

          \institute{Instituto de Astrofísica de Andalucía - CSIC, Glorieta de la Astronomía s/n, 18008 Granada, Spain\\
          \email{goswami@iaa.es}
         \and
             Observatório Nacional/MCTIC, R. Gen. José Cristino, 77, 20921-400, Rio de Janeiro, Brazil
         \and
             Scuola Normale Superiore, Piazza dei Cavalieri 7, 56126 Pisa, Italy
         \and
             Centro Astronómico Hispano en Andalucía, Observatorio de Calar Alto, Sierra de los Filabres, 04550 Gérgal, Spain\\
             }

   \date{}
\abstract
{Identifying Population III (Pop III) stars, the first generation of metal-free stars in the early Universe, remains a central challenge in astrophysics. High-ionization emission lines, such as He\,\textsc{ii}, are commonly used as tracers of Pop III signatures. However, realistic galaxies may host hybrid stellar populations, including both Pop III and metal-enriched  Population II (Pop II) stars, complicating the interpretation of observed spectra.}
{We aim to investigate how hybrid Pop III/Pop II stellar populations affect emission line diagnostics and assess the detectability of Pop III stars across different galactic environments and redshifts.}
{We select galaxies with varying Pop III-to-total  mass ratios from the \textsc{IllustrisTNG} cosmological simulations. Using self-consistent photoionization models, we compute integrated spectra by adopting local physical conditions from the simulations to study the resulting emission line diagnostics.}
{We find that emission line diagnostics are strongly dependent on the relative Pop III contribution. Current diagnostics can identify galaxies dominated totally by Pop III stars but fail for systems where Pop II stars contribute significantly, introducing degeneracies in interpreting observed spectra.}
% conclusions heading (optional)
{Our results highlight the limitations of existing emission line diagnostics in hybrid systems and emphasize the need for additional methods that account for mixed stellar populations to reliably detect Pop III stars during and after the epoch of reionization.}


   \maketitle

\section{Introduction}

The detection and characterization of the first galaxies in the universe remain one of the most compelling goals in modern astrophysics. These primordial systems, formed during the epoch of reionization, hold the key to understanding the early phases of cosmic structure formation, the initial stages of star formation, and the chemical evolution of the universe. Among the most intriguing aspects of this research is finding  Pop III stars—the first generation of metal-free stars—and their possible coexistence with subsequent generations of Pop II stars , which formed from gas enriched by the first supernovae \citep{Venditti2023,Elka2025,Venditti2026}. While Pop III stars are theorized to have been massive, short-lived, and extremely luminous, their direct detection has remained elusive, leaving their observational signatures and interplay with Pop II stars largely unconstrained  despite significant progress in infrared sensitivity and high-redshift galaxy observations with the \textit{James Webb Space Telescope (JWST)} \citep[e.g.][]{Bromm2002,Berzin2021,Curtis2023,Fujimoto2025,Roberts2024}.

A key observational signature of Pop III stars is the nebular He\,\textsc{ii} emission line, which arises from the recombination of doubly ionized helium in the presence of hard ionizing radiation ( IP > 54 eV). This line is expected to be  particularly strong in young Pop III-dominated systems due to the predicted high effective temperatures of these stars, which produce copious amounts of ionizing photons capable of doubly ionizing helium\cite[e.g.][]{Schaerer2002,Jimenez2006,Grisdale2021,Hawcroft2025}. In contrast, Pop II stars rarely produce strong He\,\textsc{ii}$\lambda$1640 emission, as only a few hot stars (e.g., early-type Wolf-Rayets, WR) can ionize He\,\textsc{ii}, making this line a powerful diagnostic for distinguishing between distant Pop III galaxies and galaxies dominated by other excitation sources. The optical He\,\textsc{ii}$\lambda$4686 nebular line has also been useful in locating hard ionizing radiation sources  in neighboring metal-poor 
 star-forming (SF) galaxies, in addition to the UV He\,\textsc{ii}$\lambda$1640 line \cite[e.g.][]{Kehrig2015,Kehrig2018,Plat2019,Ucci2019}. 
Previous studies have made significant progress in identifying high-redshift SF galaxies and analyzing their emission line diagnostics to infer stellar populations. For instance, observations of Ly$\alpha$, He\,\textsc{ii}$\lambda$1640, and [O\,\textsc{iii}] $\lambda$5007  emission lines have been used to distinguish between Pop III and Pop II dominated systems \citep{Schaerer2002,Inoue2011,Ygg2011,Sobral2015}. These diagnostics are particularly sensitive to the hardness of the ionizing spectrum, which is significantly different for Pop III stars compared to Pop II stars due to their higher effective temperatures and lack of metal line cooling \citep{Tumlinson2001,Raiter2010}. However, these studies have primarily focused on galaxies dominated by either Pop III or Pop II stars, leaving a critical gap in our understanding of systems where both populations may coexist \citep{Tornatore2007,Muratov2013,Jeon2015,Sarmento2019}. \\
The evolution of Pop III star formation over cosmic time and the possibility that these stars are not limited to the lowest-mass halos have been partially clarified by recent studies. Specifically, Pop III stars can form in more massive galaxies than previously thought, especially when metal mixing is inefficient or delayed \citep{Ciardi2005,Liu2020,Venditti2023}. 
% Recent studies have also suggested that low-metallicity star-forming regions in the local universe may retain residual Pop III-like signatures, providing an opportunity to study their spectroscopic properties in detail \citep{Senchyna2017,Kehrig2018}. Observations of extreme emission-line galaxies and blue compact dwarfs have revealed strong He,\textsc{ii} and other high-ionization lines indicative of hard radiation fields. These signatures are consistent with the presence of very metal-poor, massive stars that share key spectral characteristics with Pop III stars \citep{Berg2019,Izotov2021}.
As mentioned before, strong high-ionization lines, like He\,\textsc{ii}, have also been observed in the spectra of local, low-metallicity star-forming systems \citep[e.g.][]{
Garnett1991,Thuan2005,Senchyna2017,Berg2019,Schaerer2019,Izotov2021}. Different studies have found that standard ionizing sources cannot explain the observed spectra of such objects, and in some cases, peculiar hot (nearly) metal-free massive stars (Pop III-like) are required to reproduce the observations \citep[e.g.][]{Kehrig2015,Kehrig2018,Kehrig2021,2022Elridge,Raul2025,Mondal2025}. This might indicate that residual Pop III-like signatures can be 
present in these nearby extreme emission-line galaxies too. 

These results imply that there are probably hybrid populations of chemically evolved and (nearly) metal-free stars throughout cosmic time. Moreover, recent studies have suggested that metal enrichment in the early universe is highly inhomogeneous, with inefficient and anisotropic metal mixing allowing chemically pristine or weakly enriched gas pockets to persist even in environments that have already hosted star formation \citep{Tornatore2007,Cen2011,Jaccks2018,Skinner2020}. Comprehending this disparity is essential for constraining the circumstances that support the formation of Pop III-like stars, enhancing early star formation and chemical enrichment models, and evaluating their contribution to cosmic reionization.\\

Numerical techniques based on a variety of computational methods have been employed for studying Pop III star formation, varying by code, spatial resolution, and the implementation of feedback effects.  Adaptive mesh refinement (AMR) simulations such as \textsc{ENZO} and \textsc{RAMSES} \cite[e.g.][]{Xu2016,Pallottini2014,Skinner2020} achieve high spatial resolution and include detailed treatments of primordial chemistry and radiative feedback, typically finding very low Pop III stellar mass fractions ($\lesssim$0.1\%). Meshless finite mass and SPH simulations such as \textsc{GIZMO} and \textsc{Gadget-2} \cite[e.g.][]{Johnson2013,Maio2016,Liu2020} yield broadly consistent but somewhat higher or more variable Pop III contributions depending on feedback and mixing prescriptions. Semi-analytic models provide complementary large-volume predictions but rely on simplified gas physics \cite[e.g.][]{Visbal2020,Trinca2024}.
On the other hand, radiation-hydrodynamic simulations such as \textsc{RAMSES-RT} \cite[e.g.][]{Rosdahl2013,Skinner2020}  include more self-consistent radiative transfer, while large-volume simulations like \textsc{IllustrisTNG} \citep{Nelson2019,Pillepich2018} do not explicitly model Pop III star formation, requiring indirect identification via extremely low-metallicity stellar populations. Overall, predicted Pop III contributions vary by over an order of magnitude across models, reflecting differences in resolution, feedback, and metal mixing.

Recent JWST observations have opened a new pathway in the search for Pop~III stars. In fact, a number of galaxy candidates hosting Pop~III stellar 
populations have already been identified across various epochs, including within the Epoch of Reionization and beyond \citep{Vanzella2020,Vanzella2023,Cullen2025,Nakajima2025,
Morishita2025,Cai2025,Mondal2025}. 
Current diagnostic criteria for identifying galaxies with stellar populations dominated by  Pop III or Pop II stars rely heavily on emission line ratios, such as He\,\textsc{ii}/H$\beta$ , [O\,\textsc{iii}]/H$\beta$ and [Ne\textsc{v}]/[Ne\textsc{iii}], which are sensitive to the ionizing spectra of the stellar populations \citep{Feltre2016,Kewley2013,Hirsch2017,Nakajima2022,Chrisholm2024,Lecroq2025}.
% Furthermore, \cite{Chrisholm2024} demonstrated how the log (HeII/H$\beta$) vs. log([Ne\textsc{v}]/[Ne\textsc{iii}]) ratios evolve depending on stellar populations, IMBH, and hydrogen ionizing photons, which may also be significant diagnostics. 
However, these diagnostics often assume a chemically uniform stellar population, neglecting the potential coexistence of Pop III and Pop II stars within the same galaxy. This limitation invites further consideration of the robustness of existing criteria when applied to galaxies with mixed populations, where the ratio of Pop III to Pop II stars may vary significantly. This emphasizes the need for improved diagnostic techniques suited to differentiating mixed stellar population signatures. There have been a growing number of recent theoretical studies investigating this gap \citep{Venditti2023,Rusta2025, Venditti2026,Maiolino2026,Jeong2026}, highlighting the need for robust emission line diagnostics capable of distinguishing Pop~III signatures 
in hybrid stellar populations.
% Furthermore, the impact of nebular emission, dust attenuation, and the interstellar medium (ISM) on these diagnostics in mixed-population galaxies remains poorly understood \citep{Inoue2011,Jaskot2016,Plat2019}.
In this work, we present a proof-of-concept study of emission-line ratio diagnostics in simulated galaxies from \textsc{IllustrisTNG} simulations, explicitly including both Pop~III and Pop~II stellar populations 
in a self-consistent post-processing framework based on local cell properties. We explore how line ratios vary with the Pop~III/Pop~II mass fraction and the implications for 
standard diagnostic criteria.
We emphasize that this is not a fully predictive model of the high-redshift Universe. While \textsc{IllustrisTNG50} provides a robust and well-calibrated framework for galaxy evolution,
 it does not include dedicated Pop~III physics, and publicly available cosmological simulations that self-consistently model Pop~III star formation with comparable resolution
 and volume remain limited. We therefore adopt TNG50 as a  baseline and include an idealized treatment of Pop~III contributions in postprocessing.
Our results should thus be interpreted as illustrating qualitative behaviour and sensitivities rather than definitive quantitative predictions.
% \textbf{proof of concept stress}
% In this paper, we address this gap by exploring  emission line ratio diagnostics of simulated galaxies from cosmological simulations. Our approach complements traditional parameterized photoionization grids by explicitly modeling both Pop~III and Pop~II stellar populations within a self-consistent framework, using the local physical conditions of each cell.  We investigate how these  line ratios evolve as a function of the Pop III/Pop II mass fraction and assess the implications for current diagnostic criteria. \\
This paper is structured as follows: Section \ref{method} describes the methodology for selecting and modeling the emission line spectrum of simulated galaxies. In Section \ref{spectral} we discuss the characteristics of the integrated spectrum of the selected galaxies and then we test our galaxies on some important emission line diagnostics in Section \ref{results}. Finally, in Section \ref{conclu}, we summarize our conclusions and outlook.

\section{Methodology}
\label{method}
In this work, we use the \textsc{IllustrisTNG} simulations \citep{TNG2019,TNGP2019} to build our database of simulated observable galaxies. It is a state-of-the-art cosmological, magnetohydrodynamical simulation for galaxy formation. We use the \textsc{IllustrisTNG-50-1} box for this work, which has a cubic volume with a comoving box size of $\approx$ 51.7 Mpc per side. It generates  hundreds of unevenly spaced snapshots with a higher temporal resolution at lower redshifts to better resolve the complexities of galaxy formation and evolution processes. The redshift range covered by these snapshots is $z$ $\approx$ 20 to z = 0. 
Star formation in TNG50 occurs in cells where gas density $n_{\rm H} > 0.13,\mathrm{cm}^{-3}$, where cold high-density gas is placed
on an equation of state between temperature and density \citep{Springel2003}. Each star particle represents a single stellar population (SSP) with a \cite{Chabrier2003} Initial Mass Function (IMF) from $0.1$ and $100,M_\odot$. Stellar evolution and chemical enrichment are followed self-consistently, tracing nine individual elements (H, He, C, N, O, Ne, Mg, Si, Fe).
Feedback from stars includes (i) Type II/Ia supernovae, (ii) AGB stars, and (iii) neutron-star (NS)–NS mergers. Galactic winds driven by supernova feedback are also included in the TNG50 simulations \citep{Pillepich2018}. Black hole seeding, growth, and AGN feedback are also included \citep{Weinberger2017}. \\

It is important to emphasize that \texttt{IllustrisTNG50} \citep{Pillepich2018, Nelson2019} does not include a dedicated treatment of Pop~III star formation or 
feedback. In particular, the simulation adopts a universal Chabrier IMF and therefore does not capture the top-heavy IMF commonly expected for Pop~III stars 
\citep{Bromm2004,Hirano2014}. Similarly, metal enrichment and mixing are implemented through subgrid prescriptions calibrated primarily for galaxy evolution at 
lower redshifts \citep{Vogelsberger2013,Weinberger2017}, which may affect the predicted metallicities of subsequent stellar generations and the inferred values of 
$M_{*,{\rm PopIII}}/M_{*,{\rm total}}$. In addition, unresolved metal transport can produce artificially pristine gas pockets \citep{Jaccks2018,Skinner2020}, potentially 
biasing the identification of Pop~III-forming regions.

More specialized simulations including explicit Pop~III prescriptions and radiative feedback have recently been developed, such as the \texttt{dustyGadget} hydrodynamical simulations 
\citep{Graziani2020}, the \textsc{Thesan} radiation-hydrodynamics framework \citep{Kannan2022}, the Renaissance simulations \citep{Xu2016}, the FirstLight simulations 
\citep{Ceverino2017}, and the NEFERTITI simulations targeting primordial star formation and metal enrichment in the early Universe \citep{Koutsouridou2023}. These 
studies generally include more detailed treatments of primordial chemistry, Pop~III star formation criteria, radiative feedback, and inhomogeneous enrichment. However, 
large-volume cosmological simulations that simultaneously provide publicly available galaxy catalogs, sufficient spatial resolution, and self-consistent Pop~III physics 
remain relatively limited. For this reason, we adopt TNG50 as a practical and well-characterized framework in which Pop~III-like stellar populations can be explored 
through post-processing.

Accordingly, Pop III/ hybrid stellar populations identified in our analysis should be interpreted primarily as proxies for Pop~III-like stellar populations 
rather than as fully self-consistent primordial stars. Our goal is therefore not to provide definitive predictions for Pop~III galaxy formation, but rather a proof-of-concept 
investigation of how localized Pop~III-like components may affect nebular emission-line diagnostics in mixed stellar populations. Further discussion of these limitations is 
provided in Sec.~\ref{caveats}.
We selected a few simulated galaxies that could be ideal candidates to host Pop III populations. Galaxies and haloes in TNG50 are identified using the Friends-of- Friends (FoF) and SUBFIND substructure-identification algorithms (see \cite{Davis1985} ,\cite{Springel2001} respectively). The selection criteria for the galaxies analyzed in this study are detailed below. 
\begin{table}
\small
\centering
\resizebox{\columnwidth}{!}{
\begin{tabular}{l c c c c c}
\toprule
Name & $z$ & $\log(M_{\rm tot}/M_\odot)$ & $m_U$ & $R$ (pc) & $M_{\rm PopIII}/M_{\rm tot}$ \\
\midrule
SG1 & 10 & 7.40  & $-11.92$ & 130.1 & 1.0 \\
SG2 & 6  & 8.15  & $-8.88$  & 138.4 & 1.0 \\
SG3 & 6  & 7.98  & $-12.24$ & 106.1 & 0.8 \\
SG4 & 6  & 8.55  & $-11.43$ & 166.3 & 0.5 \\
SG5 & 0  & 10.34 & $-14.59$ & 97.8  & 0.1 \\
\bottomrule
\end{tabular}
}
\caption{Simulated galaxy properties: redshift, total stellar mass, U-band magnitude, cell size, and the ratio $M_{*,{\rm PopIII}}/M_{*,{\rm total}}$.}
\label{tab:galaxy_data}
\end{table}
\subsection{Selection criteria}
The selection criteria for the simulated galaxies from the \textsc{IllustrisTNG} simulations are based on the methodology employed in \cite{Grisdale2021}, which is summarized here as follows :
\begin{itemize}
    \item In order to investigate the distinctions between galaxies that are entirely dominated by PopIII stars and those that may have  hybrid stellar populations (Pop III + Pop II stars), we narrow down the sample of galaxies by taking into account those that have $M_{*,PopIII}/M_{*,total} = 1.0,0.8,0.5, 0.1$. \\ 
Here, $M_{*,PopIII}$ and $M_{*,total}$ represent the total mass of Pop III stars and the galaxy's total stellar mass, respectively. Stellar mass refers to the total bound stellar mass of the Subhalo (SubhaloMassType[4] in TNG50), including all stellar particles.
\item We used the metallicity of the star particles in a galaxy to assess if it is Pop III or Pop II.
We define a star particle as a Pop III one if its initial metallicity satisfies $Z_{\mathrm{init}} < Z < Z_{\mathrm{crit}}$, where $Z_{\mathrm{crit}} = 0.02\, Z_{\odot}^2 (= 4 \times 10^{-4} Z_{\odot}$ ) \citep{Ygg2011, Bromm2002, Schaerer2002}.

\item  The stellar half-mass radius for the Pop III stars
$(R_{0.5,P3})$ should be less than one kiloparsec for a galaxy to be considered \citep{Grisdale2021}.
\end{itemize}

In this study, we adopted $Z_{\mathrm{crit}} = 0.02\, Z_{\odot}^2 = 4 \times 10^{-4}Z_{\odot}$ , motivated by the implementation of population synthesis in \textsc{Yggdrasil} \citep{Ygg2011} and by various other theoretical studies of the Pop~III–Pop~II transition (e.g. \cite{Bromm2001,Bromm2003,Schneider2012}). Although the literature covers a wider range of possible $Z_{\mathrm{crit}}$ values (from $10^{-6}$ to, $10^{-2} Z_\odot$ depending on whether dust cooling, fine structure line cooling, or turbulence is assumed to set the transition \citep{Omukai2005,Schneider2006,Maio2010}, our choice reflects a conservative limit. It should be noted that the choice of $Z_{\mathrm{crit}}$  can affect both the predicted Pop~III fraction and the resulting emission line signatures. A lower $Z_{\mathrm{crit}}$ would suppress the Pop~III contribution, reducing the number of galaxies in which Pop~III-like emission-line signatures are detectable, and shifting galaxies more rapidly into the Pop~II-dominated regime. On the other hand, a higher $Z_{\mathrm{crit}}$ value will lead to an extended period during which galaxies can exist either in Pop~III or mixed-dominance. In this sense, $Z_{\mathrm{crit}}$ primarily regulates the relative abundance of detectable Pop~III-influenced galaxies in the simulations.\\
We impose $R_{0.5, \mathrm{P3}} < 1\,\mathrm{kpc}$ for the stellar half-mass radius of Pop~III star particles. This constraint ensures that Pop~III stellar systems remain physically compact, as expected for their formation in low-mass minihaloes at $z \gtrsim 10$. Cosmological simulations consistently predict that the earliest star-forming systems occupy very small physical volumes, with stellar half-mass (or half-light) radii typically of order tens to a few hundred parsecs (e.g. \citealt{Ma2017, Milosavljevic2015, Griffen2016, Hyung2017}). Similar compactness criteria have also been adopted in other high-resolution simulations (e.g. \citealt{Grisdale2021}) to avoid unrealistically extended Pop~III star particles. 
\begin{figure}
    \centering
    \includegraphics[width=0.5\textwidth]{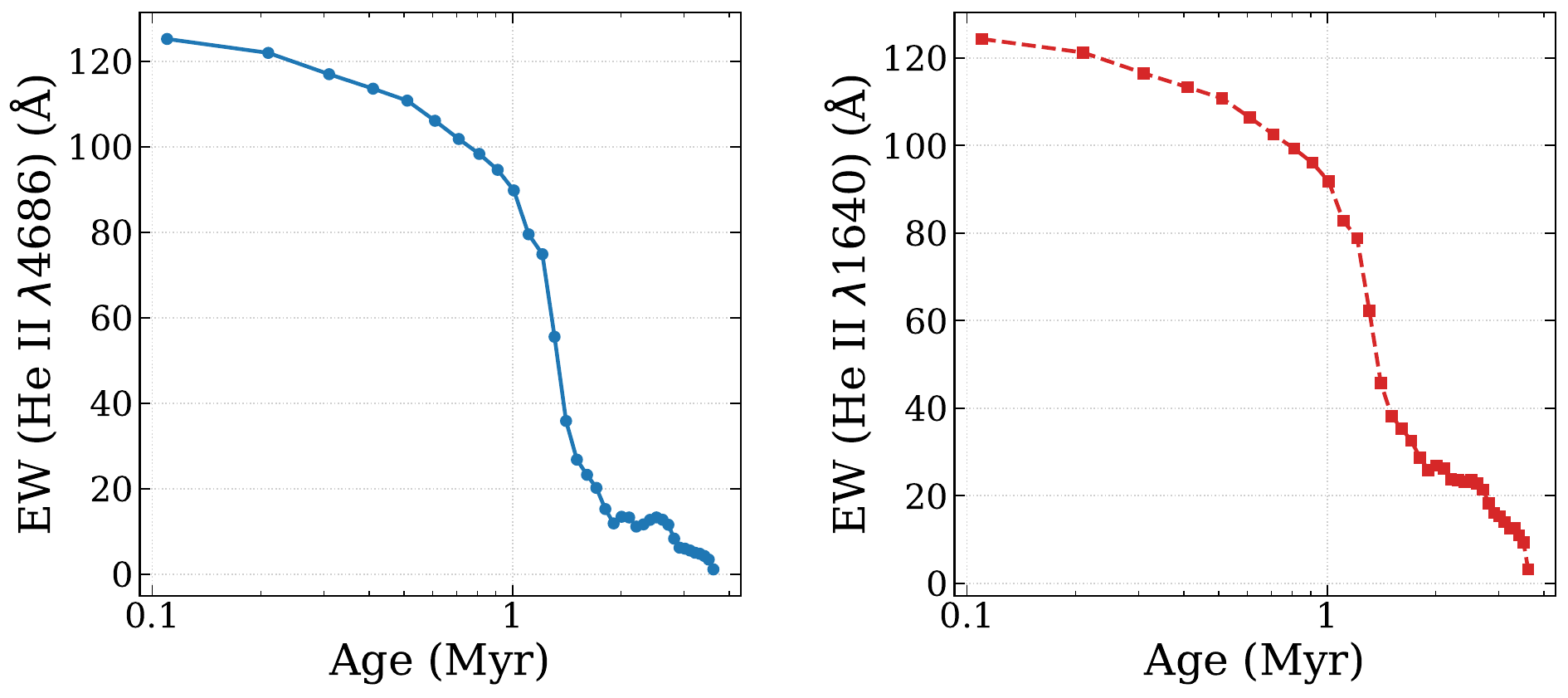}
    \caption{ Evolution of the EW of the He II emission lines as a function of stellar population age for Population III models. The left panel shows the EW of the nebular He II$\lambda$4686  line, while the right panel presents the EW of the HeII$\lambda$1640  line.}
    \label{fig:EWAge}
\end{figure}
Physically, this upper limit is supported by the expected size of the corresponding \textit{Strömgren sphere} which we discuss in Sect.\ref{cloudy}.
Since the strength of the He\,\textsc{ii} emission varies inversely with the age of very massive stars, therefore to ensure the presence of massive, short-lived stars ( $\lesssim 3$ Myr) capable of producing stronger He\,\textsc{ii}  emission, we select the youngest galaxies for each redshift case. The ages of individual star particles are computed from their formation times (\texttt{GFM\_StellarFormationTime} in TNG50 simulations data products, referring to the exact time the star was formed), and these are then converted to lookback times using the \texttt{astropy.cosmology} framework. This method ensures that our sample preferentially includes systems whose stellar populations are dominated by recently formed very young stars i.e. typically $\lesssim$3 Myr  and reduces the amount of evolved or  "dead" Pop III-like stellar particles that no longer make a substantial contribution to the ionizing radiation field. On the other hand, the gas ionized by the young populations is likely to show strong nebular He II emission because very massive, short-lived stars emit the largest number of ionizing photons, as we have discussed earlier. To test this dependency of age on the nebular He II emission, we run photoionization models using spectral energy distribtions (SEDs) generated with the \textsc{Yggdrasil} synthesis code, of burst of stellar populations of different ages, while keeping all other variable parameters (as mentioned in Sect.~\ref{cloudy}) constant. This dependency is clearly shown in Fig. \ref{fig:EWAge}, where both the He II$\lambda$4686  and He II$\lambda$1640  EWs show a steep decline with increasing age. The plots demonstrate that He II emission is strongest at the earliest stages of evolution and rapidly weakens as the stellar population evolves to older ages.\\
\begin{figure*}% 't' places the figure at the top of the page
    \centering
    \includegraphics[width=0.18\linewidth]{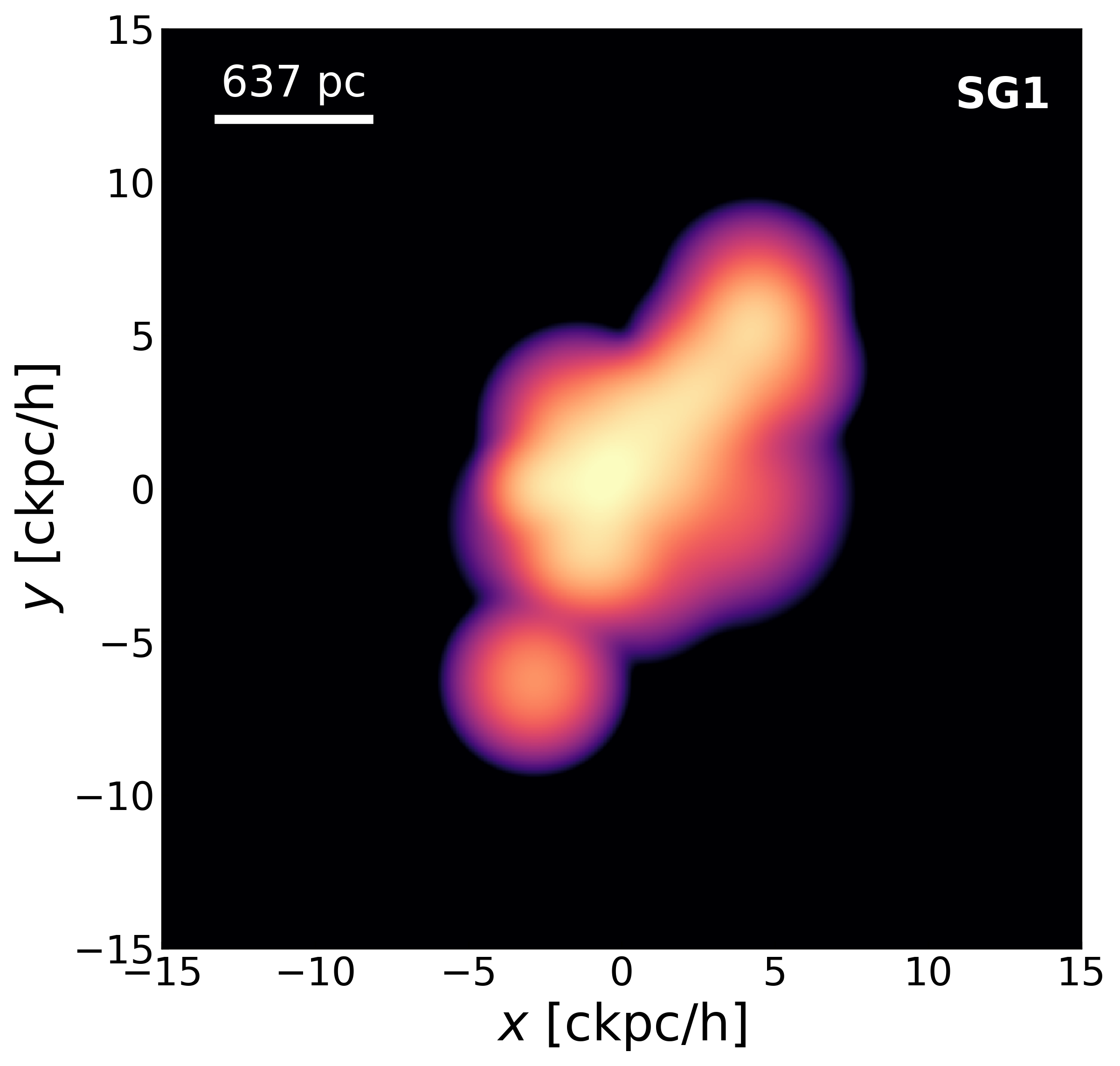}
    \includegraphics[width=0.18\linewidth]{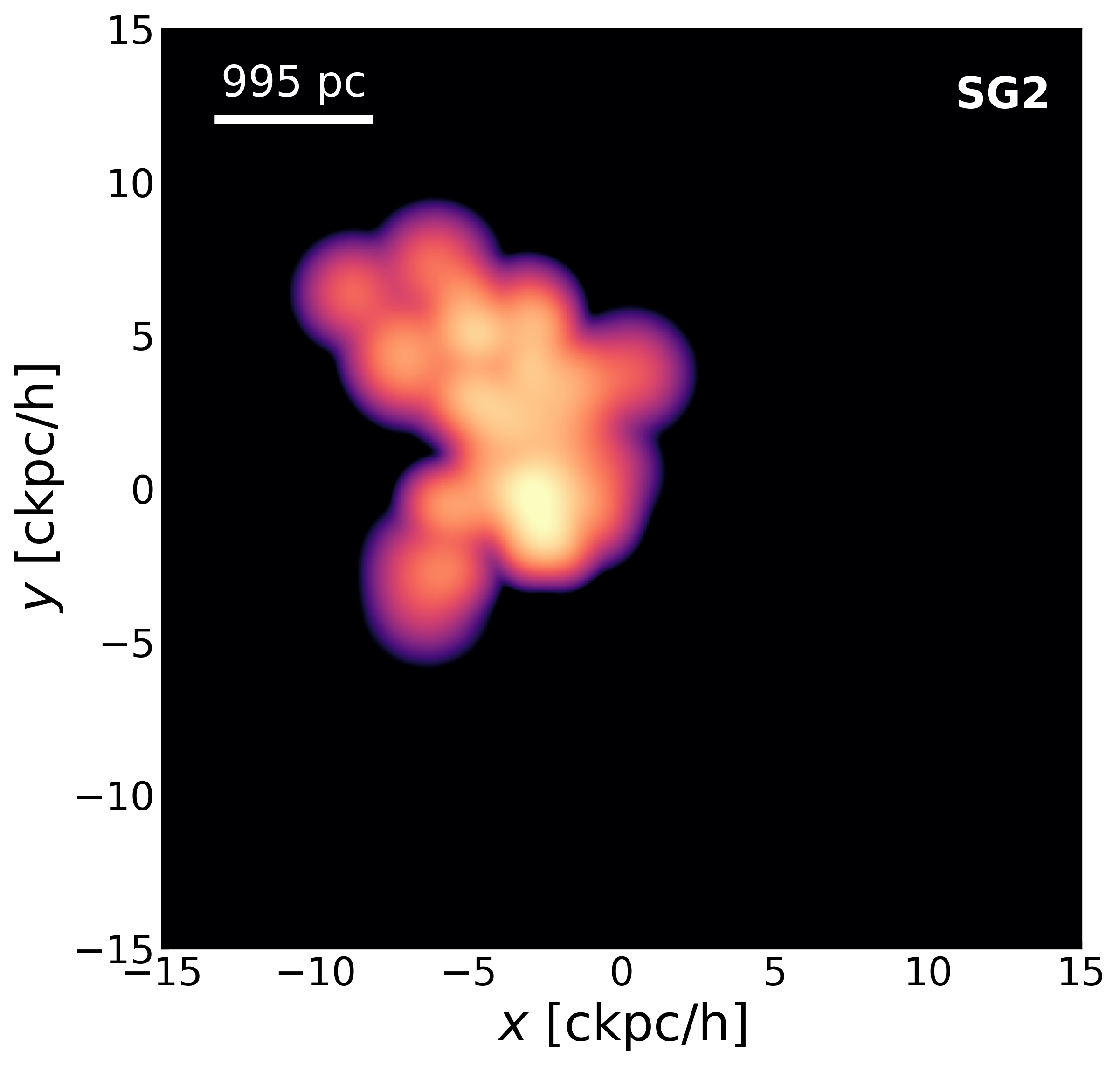}
    \includegraphics[width=0.18\linewidth]{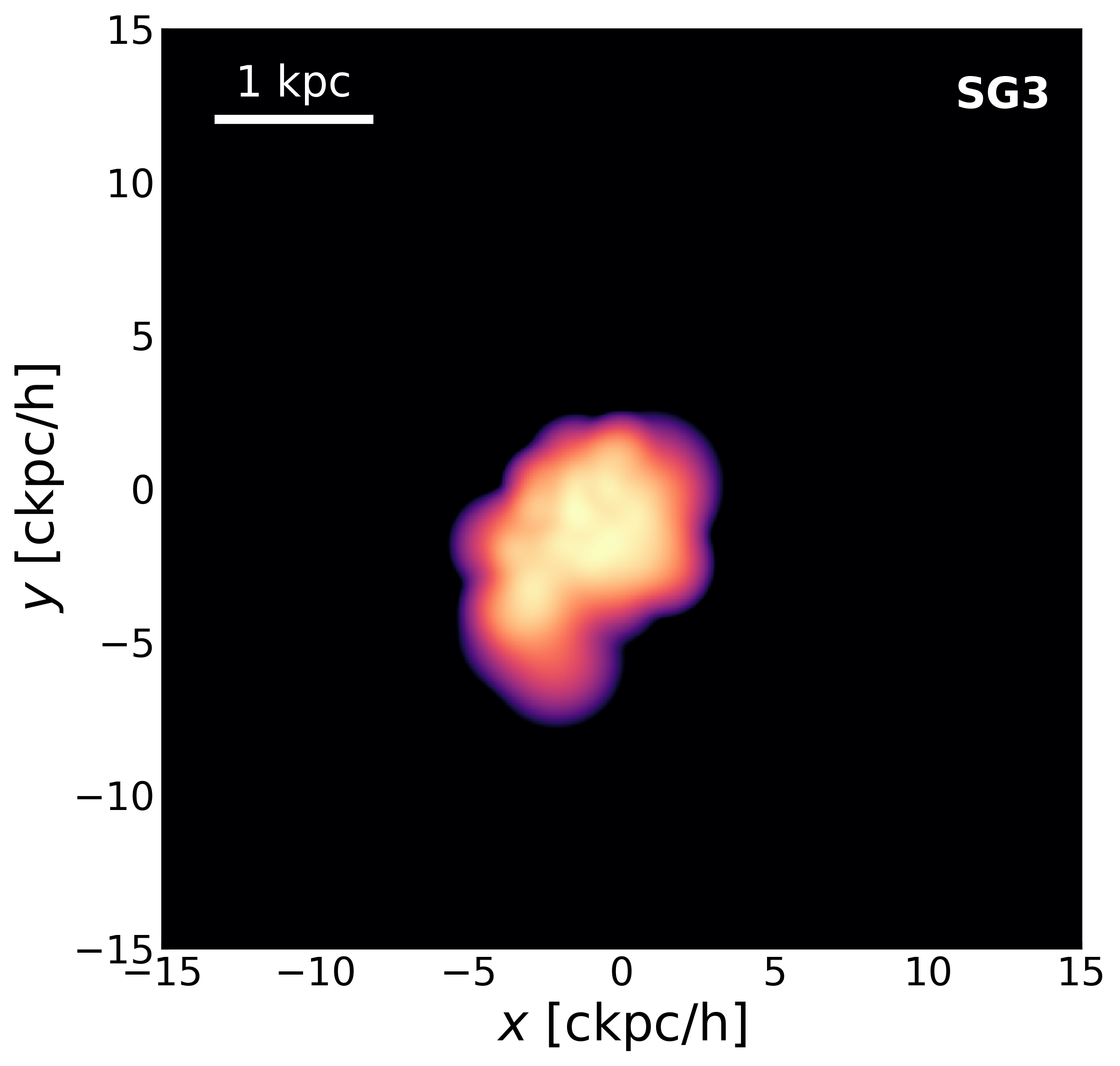}
    \includegraphics[width=0.18\linewidth]{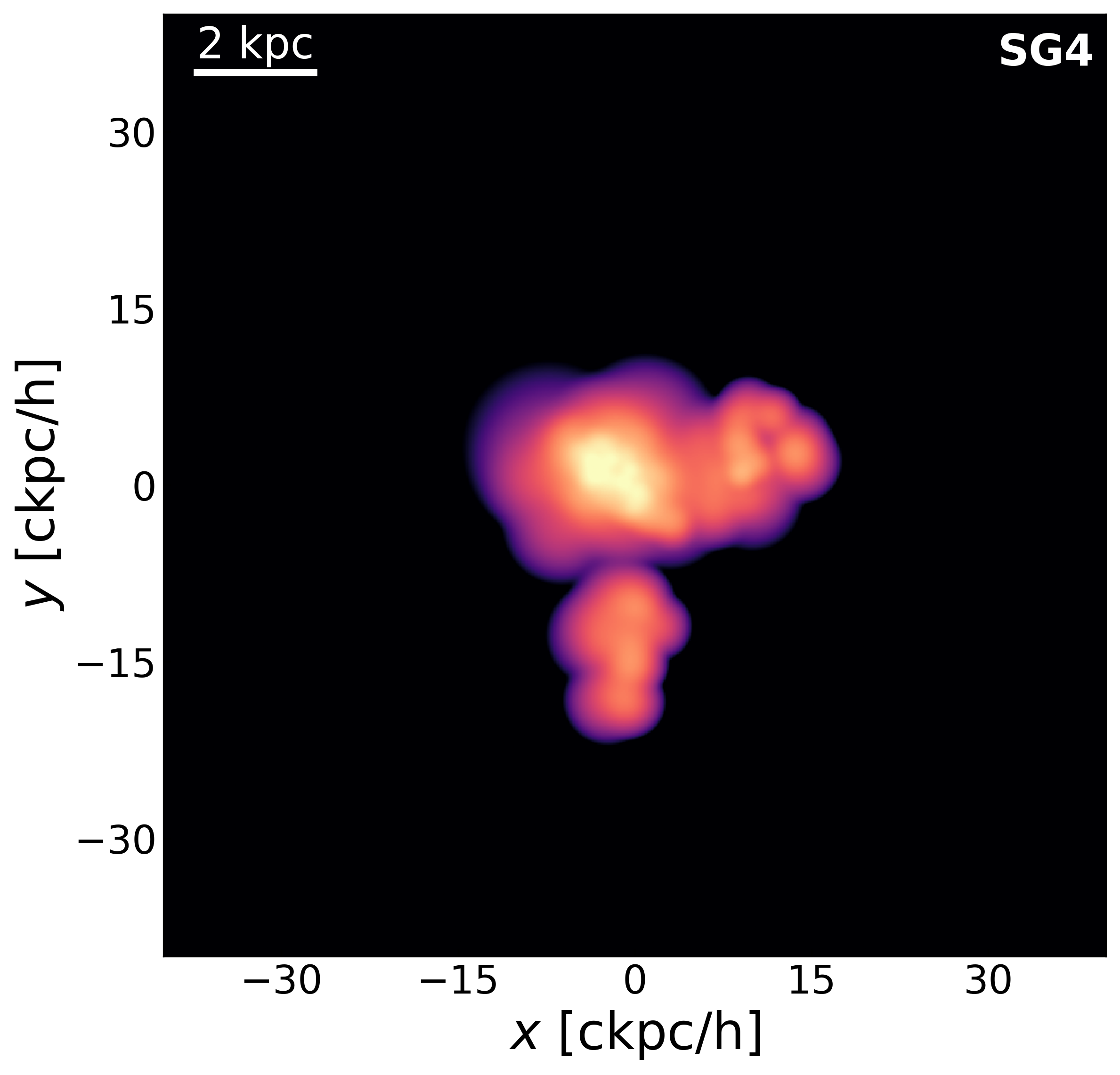}
    \includegraphics[width=0.21\linewidth]{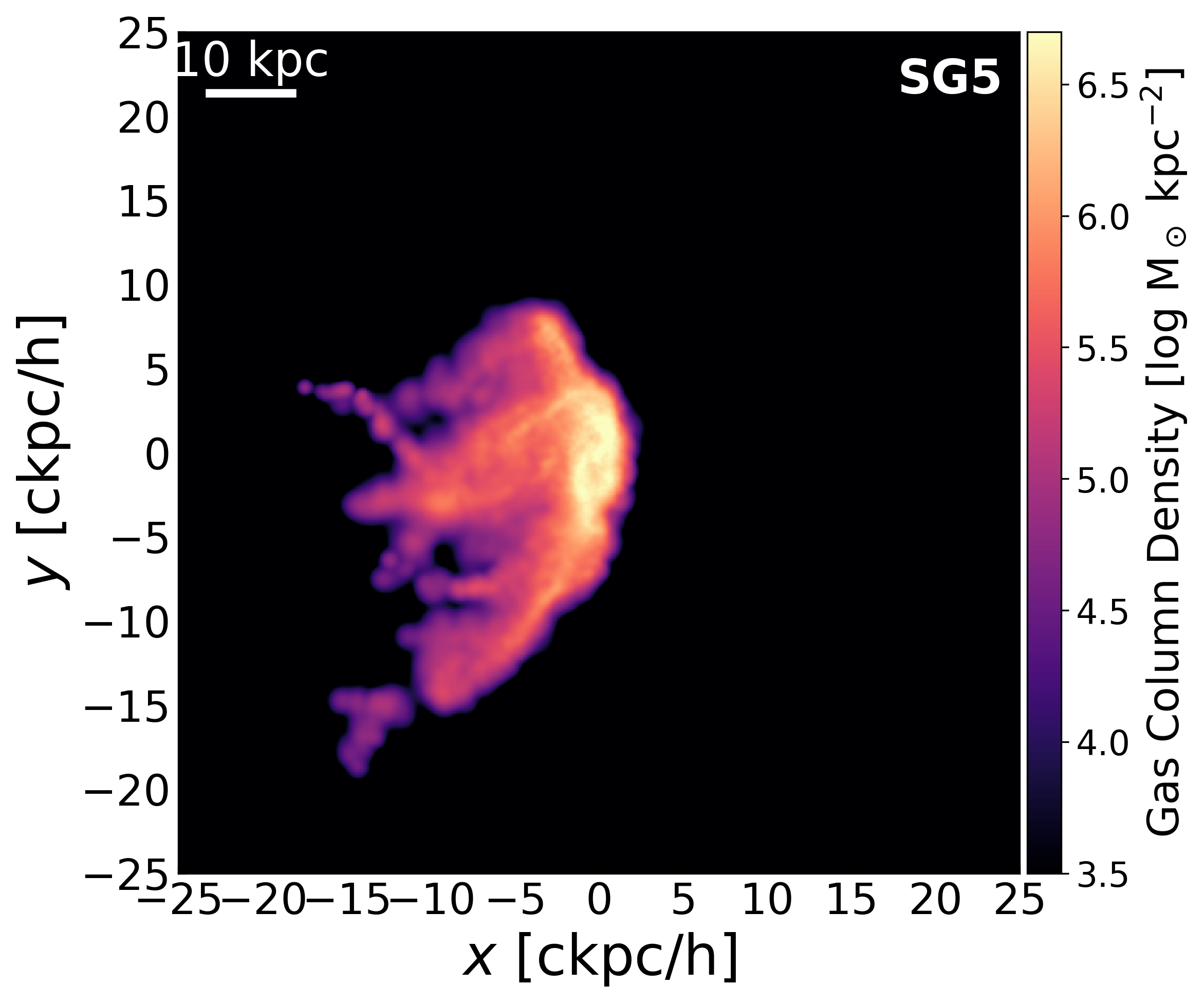}
    \caption{Gas column density maps of the five galaxies selected, SG1, SG2, SG3, SG4, and SG5 (from left to right, respectively), from the TNG-50-1 simulations. The gas column density values are displayed in the color bar on the right, while the physical scales are displayed at the upper left of each galaxy.}
    \label{fig:TNG50}
\end{figure*}
Table \ref{tab:galaxy_data} provides a list of the galaxies selected, their characteristics, and the ratio of $M_{*,PopIII}/M_{*,total}$. Figure \ref{fig:TNG50} shows the gas column density maps of such objects from the TNG50 simulations. Our sample consists of five galaxies, chosen to study Pop III-dominated and mixed Pop III/II systems across cosmic time. At $z$=6, we select three galaxies with different Pop III fractions ($M_{*,PopIII}$/$M_{*,total}$), targeting systems where hybrid populations can explored. To explore redshift evolution, we include one young galaxy at 
$z$=10 and one at $z$=0. Selecting the youngest galaxies in ease case ensures we identify galaxies whose stellar populations are dominated by recently formed stars,
which are expected to produce the hardest ionizing spectra and strongest nebular He II emission. 
We select most galaxies at $z=6$ to investigate the properties of galaxies around the epoch of reionization, a crucial period for understanding early and very metal-poor stellar populations and their imprint on the surrounding interstellar medium. 
At this redshift, Pop~III systems that satisfy our criteria are extremely rare in TNG50-1, representing only $\sim10^{-5}$--$10^{-4}\%$ of all galaxies. 
Among these, SG3 (see Table~\ref{tab:galaxy_data}) has a Pop~III stellar mass fraction of $M_{*,\mathrm{PopIII}}/M_{*,\mathrm{total}} = 0.8$, occurring in only $\sim1.1\times10^{-5}\%$ of all galaxies at $z=6$, making it one of the most extreme partially metal-free systems in the simulation. 
In contrast, SG4, with a lower Pop~III mass fraction, appears more frequently ($\sim10^{-4}\%$), providing a complementary example of a less extreme Pop~III galaxy at the same epoch. 
To probe redshift evolution, we also include SG1, a fully Pop~III-dominated system ($M_{*,\mathrm{PopIII}}/M_{*,\mathrm{total}}=1$) at $z=10$, where such pristine systems are slightly more common ($\sim2\times10^{-4}\%$). 
Finally, to explore the presence of possible analogues in the local Universe, we include SG5 at $z=0$, which exhibits an upper-limit Pop~III fraction of $M_{*,\mathrm{PopIII}}/M_{*,\mathrm{total}}=0.1$, found in only $\sim5\times10^{-5}\%$ of galaxies at this redshift. 
This rarity indicates that systems that host Pop~III stars are effectively extremley rare in the nearby Universe within TNG50. Interestingly, observations of extremely metal-poor dwarf galaxies  at z=0 as I Zw 18  and SBS0335–052E have revealed very hard and puzzling ionizing spectra with bright nebular He II emission, suggesting these galaxies may host peculiar very hot massive stars i.e.  PopIII-like stars; \citep[see e.g.,][]{Thuan2005,Kehrig2015,Kehrig2018,Perez2020,Arroyo2025,Mingozzi2025,Szeci2025}. These works have revealed unexpectedly hard ionizing spectra and nebular He II emission, suggesting these galaxies may host either Pop III or very metal-poor, massive stellar populations. SG5 thus serves as a limiting case for investigating how residual PopIII-like populations might influence emission-line diagnostics.
\begin{figure*}
    \centering
    \includegraphics[width=0.35\textwidth]{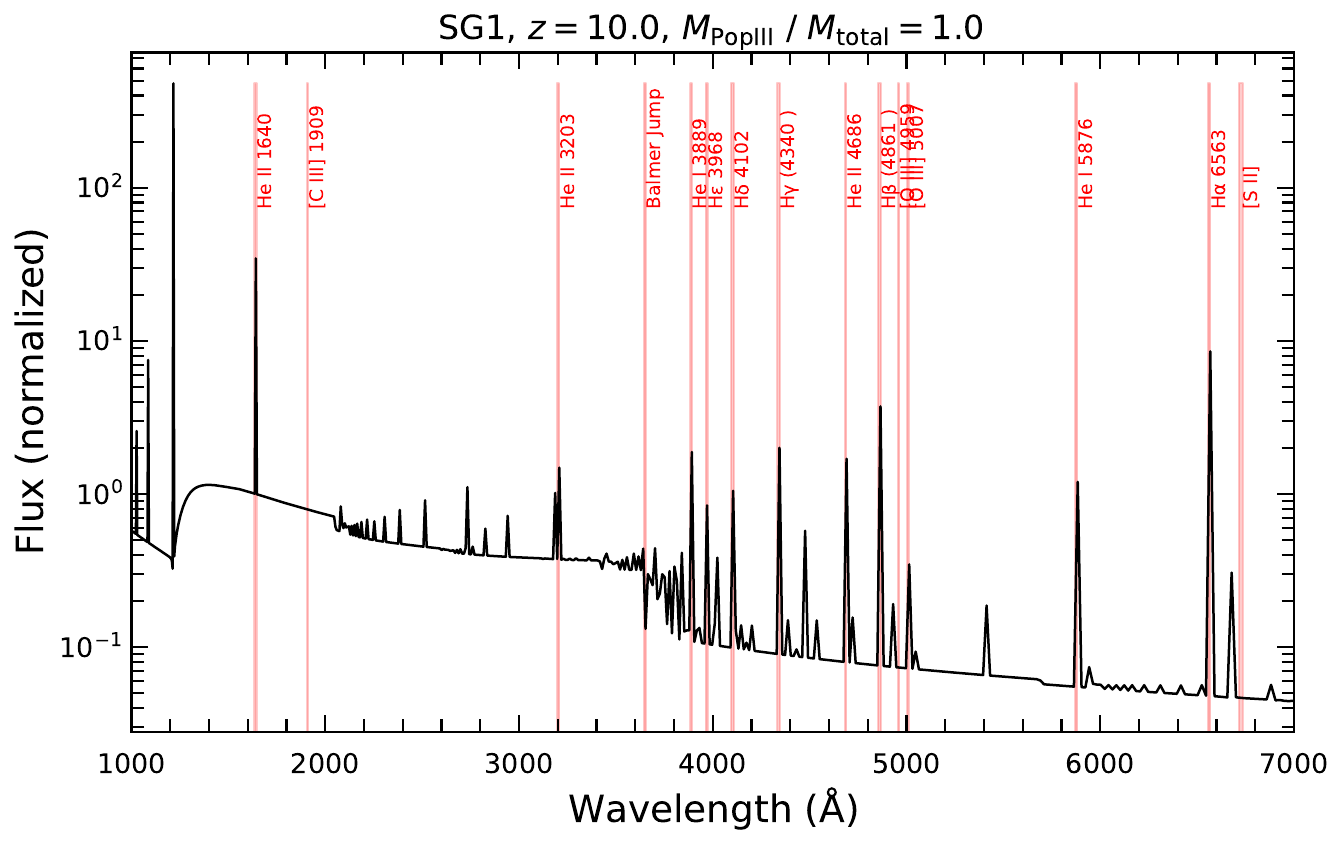}
    \includegraphics[width=0.35\textwidth]{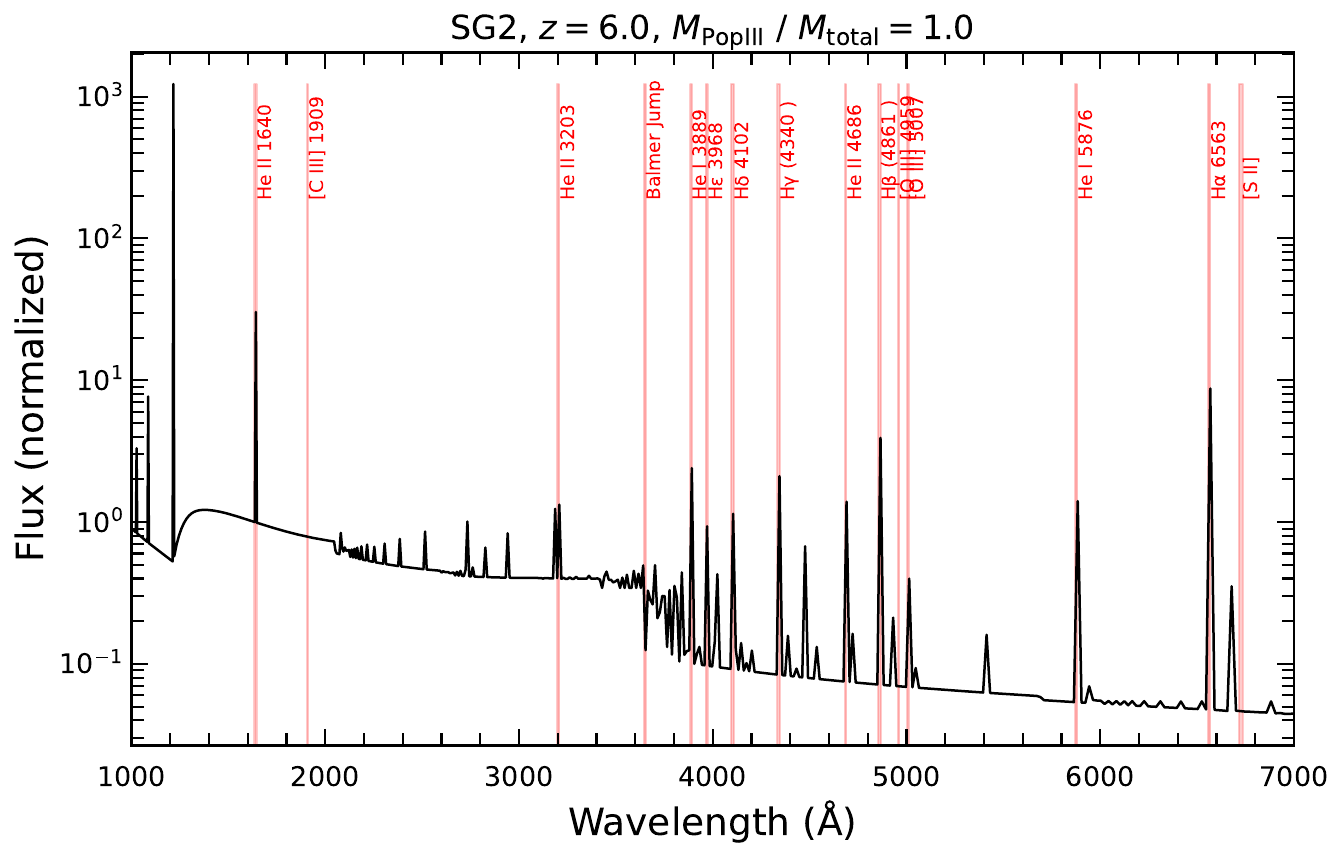}
     \includegraphics[width=0.35\textwidth]{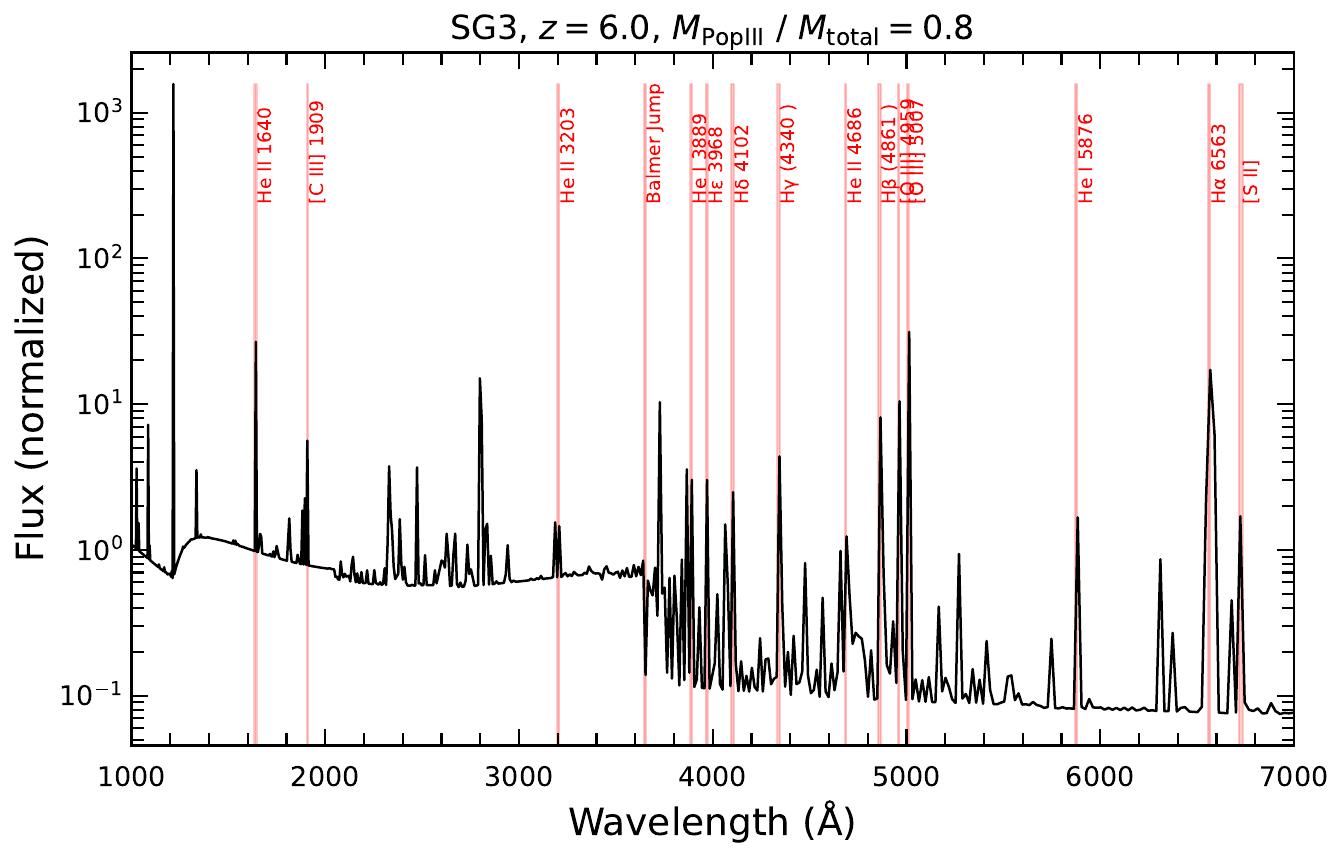}
     \includegraphics[width=0.35\textwidth]{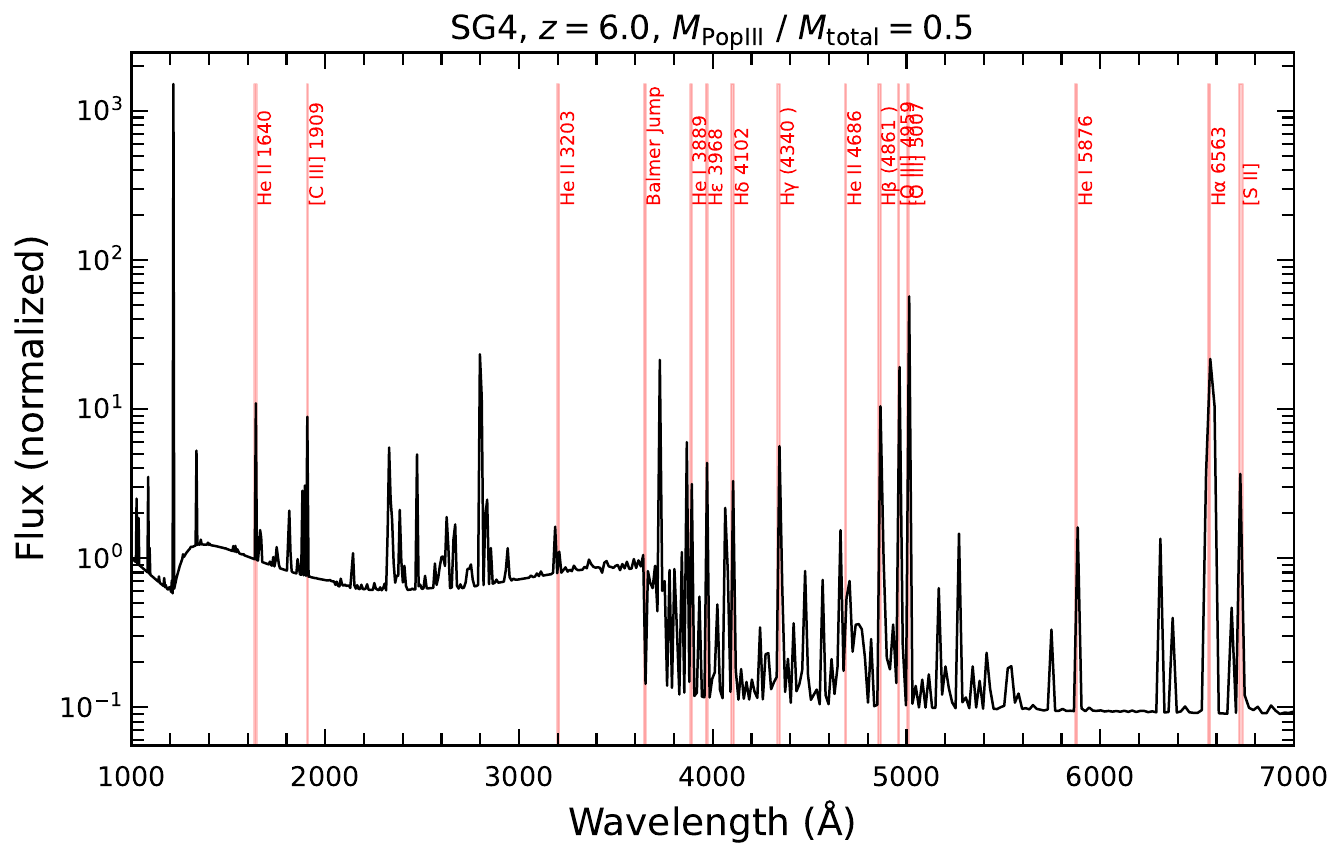}
     \includegraphics[width=0.35\textwidth]{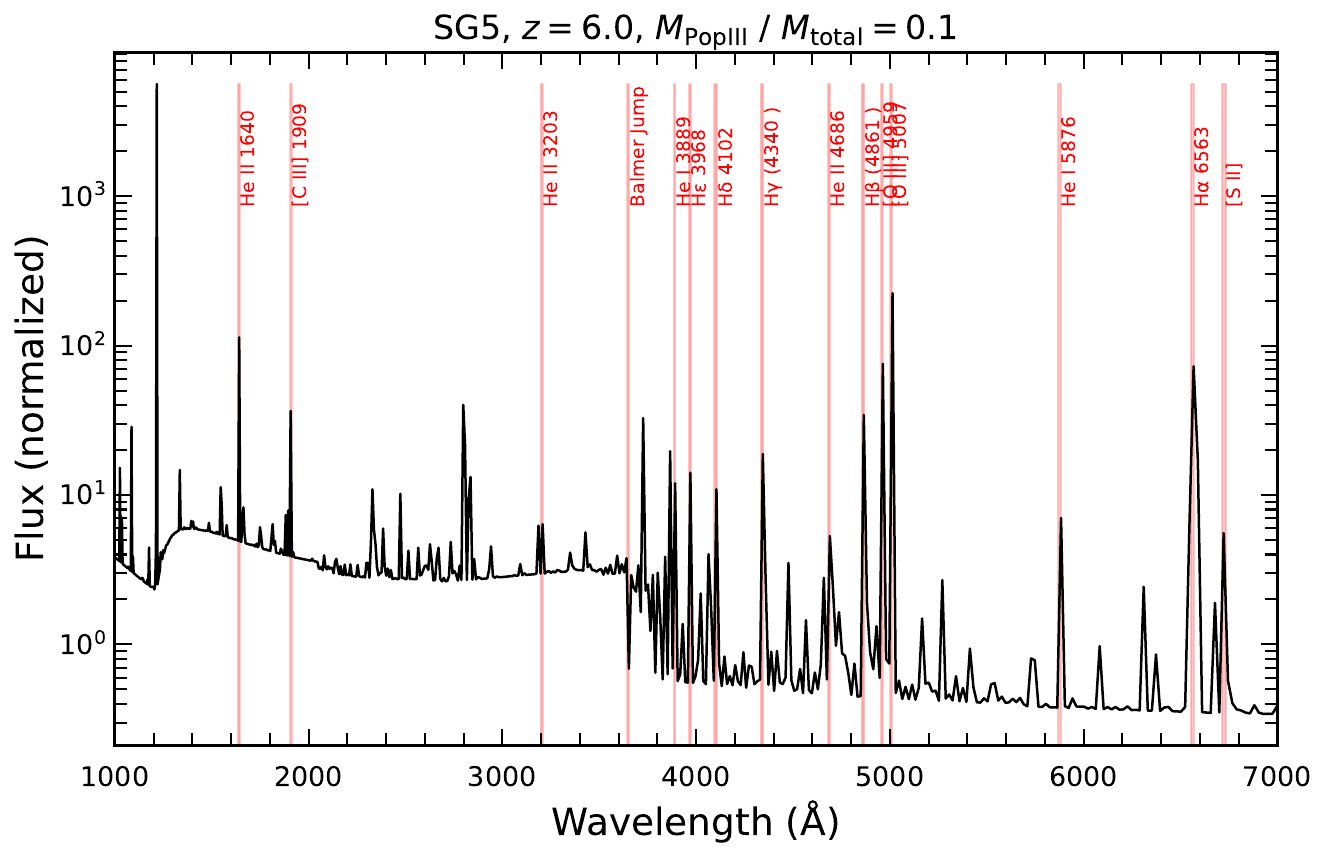}
    \caption{Integrated spectrum of the five simulated galaxies in this work post-\textsc{Cloudy} asssuming a PopIII.1 IMF and a spherical geometry.  All the spectra are normalized to the flux at 1500 \AA{},  Flux$_{1500}$ . Some key emission lines to distinguish between PopIII and PopII characteristics are marked in red shaded regions.  }
    \label{fig:intspe}
\end{figure*}

% We select most galaxies at $z$=6 to investigate the properties of galaxies around the epoch of reionization, a crucial period for understanding the properties of early galaxies. We also add a 100\% PopIII-dominated system, SG1, at a higher redshift to check the effect of varying $z$ on the emission line diagnostics.  Additionally, motivated by the presence of Pop III candidate signatures in more massive local galaxies discussed in the previous section, we select SG5 at $z$=0. In the TNG-50-1 simulations, we do not find galaxies that have the ratio  $M_{*,PopIII}/M_{*,total}$ above 0.1  at this redshift, indicating that more extreme Pop III-dominated systems are not produced in TNG50. We use the case of galaxy SG5, with this upper limit ratio of  $M_{*,PopIII}/M_{*,total}$ =  0.1 to study the emission line properties of nearby galaxies possibly hosting Pop III-like candidates. 
% For each one of the selected galaxies from TNG-50-1 simulations we can now derive its corresponding spectrum. To do so, first we have made use of the spectral synthesis code \textsc{Yggdrasil} \citep{Ygg2011} to derive the stellar SEDs of the galaxies. Once we have the ionizing SEDs then we can produce their spectra after running the \textsc{Cloudy} photoionization code .

In the TNG simulations, each stellar particle is assigned a formation age and metallicity. We associate each particle with the corresponding SED from \textsc{Yggdrasil} based on its age and classify it as either Pop~III or Pop~II according to its metallicity. The post-processing approach allows us to explore all three Pop~III IMFs implemented in \textsc{Yggdrasil} \citep{Ygg2011} independently. The resulting SEDs at different stellar ages are shown in Fig.~\ref{fig:grasil}.

The first and our primary IMF corresponds to the Pop~III.1 case and assumes an extremely top-heavy stellar mass distribution described by a Salpeter power law,
\begin{equation}
\phi(M) \propto
\begin{cases}
M^{-\beta}, & 50 \leq M/M_{\odot} \leq 500, \\
0, & \text{otherwise},
\end{cases}
\label{eq:imf_popiii1}
\end{equation}
where $\beta = 2.35$ \citep{Salpeter1955}. 

The second IMF is the Pop~III.2 case and it is  moderately top-heavy. It is modeled as a log-normal distribution,
\begin{equation}
\phi(M) \propto
\frac{1}{M}
\exp\left[
-\frac{(\ln M - \ln M_{\mathrm{c}})^2}{2\sigma^2}
\right],
\quad 1 \leq M/M_{\odot} \leq 500,
\label{eq:imf_popiii2}
\end{equation}
with characteristic mass $M_{\mathrm{c}} = 10\,M_{\odot}$ and dispersion $\sigma = 1.0 M_{\odot} $ \citep{Tumlinson2006,Raiter2010}.

As a comparison case, we also consider a low-mass-dominated Pop~III IMF with the same functional form as the Kroupa IMF,
\begin{equation}
\phi(M) \propto
\begin{cases}
M^{-1.3}, & 1 \leq M/M_{\odot} < 0.5, \\
M^{-2.3}, & 0.5 \leq M/M_{\odot} \leq 100,
\end{cases}
\label{eq:imf_kroupa}
\end{equation}
following \citet{Kroupa2001}. Although such an IMF is not strongly supported by current theoretical models of primordial star formation, it provides a useful reference case.

By exploring all three IMFs, we bracket the theoretical uncertainty in the characteristic masses of Pop~III stars and assess the sensitivity of our results to the assumed primordial stellar mass distribution.

For particles with $Z > $ $Z_{crit}$, we employ the PopII.K (Z = 0.02, Kroupa IMF \citep{Kroupa2001}) provided by the \textsc{Yggdrasil} models. To model the SED of a star particle,  we select the SED from \textsc{Yggdrasil} that matches both the age of the star particle ($t_{\star,\text{age}}$) and the appropriate IMF, based on its metallicity. In combination with the IMF, \textsc{Yggdrasil} employs the birth mass of the star population, $M_{\text{SED}}$, to calculate the SED's shape and magnitude.

Every star particle in TNG-50-1 has an assigned initial mass, $M_{\star,\text{initial}}$. However, TNG50 does not include Pop~III star formation or evolution prescriptions specifically, therefore, we do not use $M_{\star,\mathrm{initial}}$ directly. Instead, in post-processing, we reinterpret each selected star particle representing a Pop III stellar population, and so we assign it a Pop~III IMF within \textsc{Yggdrasil}. The template SED produced by \textsc{Yggdrasil} corresponds to a fixed stellar mass, $M_{\text{SED}}$. To ensure consistency, we scale the luminosity of the SED of \textsc{Yggdrasil} in order to pass a suitable value of $L_{\text{tot}}$ to \textsc{Cloudy}. This is calculated using
\begin{equation}
L_{\text{tot}} =
\frac{M_{\star,\text{initial}}}{M_{\text{SED}}}
\int L_{\lambda}\,\mathrm{d}\lambda .
\label{eq:Ltot}
\end{equation}

where $L_{\lambda}$ is the luminosity of the SED at each wavelength ($\lambda$).
$M_{\star,\text{initial}}$ therefore scales the magnitude of the SED, while the
choice of IMF and $t_{\star,\text{age}}$ determines its shape for each particle.

\subsection{Constructing spectra with \textsc{Cloudy}}
\label{cloudy}

\textsc{Cloudy} \citep{Cloudy2017} models are then run for each star particle using its  \textsc{Yggdrasil} SED,  and the total luminosity ($L_{\text{tot}}$) and hydrogen number density ($n_{H}$) of its
host cell as inputs. \textsc{Cloudy} photoionization code is used to determine the strength of the emission lines, and it is calculated till the radius of $R$/2 where $R$, the host cell size, is provided in the TNG-50-1 galaxy cutouts. For star particles with a Pop III IMF, we assume constant-density gas clouds, no dust, and  primordial abundance ( 76\% of the gas  is composed of hydrogen and the remainder of the gas is composed of helium). However, for Pop II particles, heavier elements can be present since they are not expected to have primordial element abundances dominated by H and He. In our \textsc{Cloudy} models, for Pop III particles, the gas abundances are explicitly specified to represent primordial abundances with only H and He as mentioned above, while all other elements are disabled. For Pop II star particles instead, gas-phase abundances are not explicitly defined in the \textsc{Cloudy} input where these models are run with \textsc{Cloudy} default abundance setup and the metallicity dependence on the ionizing spectrum is incorporated using the \textsc{Yggdrasil} input SED for which we adopt Z=0.02. Thus, our approach should be regarded as an idealized nebular modelling framework
aimed primarily at capturing the ionizing properties and resulting HeII signatures of young stellar populations, rather than a fully self-consistent chemo-dynamical 
treatment of the ISM. We further explored whether the choice of gas geometry in \textsc{Cloudy} affects the detectability of Pop~III stars, adopting both spherical and plane-parallel geometries.
In addition to the radiation field from stars,
we include a background radiation field consistent with \cite{Cloudy2017}. This field is designed to
mimic the observed cosmic radio to X-ray background with contributions
from the CMB, assumed to be a black body with a temperature
of $T_{CMB}$ = 2.725(1 + $z$)K.\\
The cube of a selected galaxy is divided into host cells of equal sizes, and 
if a host cell has one or more star particles, a spectrum is calculated for
each particle, which is then summed to create a single integrated spectrum for
that cell.   Finally, the spectrum of each cell is stacked to form the integrated spectrum of the galaxy.  We note here that since we select the youngest galaxy at each reshift, we do not apply an age cut for stellar particles in our fiducial analysis in reconstructing the final spectrum. Nevertheless, we investigate the effect of the commonly adopted age cut of 10 Myr \citep{Westera2004,Xiao2018} for nebular emission calculations, while treating older particles as "dead" hence not contributing to the final spectrum, and we discuss its impact on the emission line diagnostics in Sect.~\ref{results}. .This procedure is carried out for the five
galaxies listed in Table \ref{tab:galaxy_data} in order to compare and examine the spectral characteristics which are discussed in the following section.
Regarding the stellar half-mass radius criterion for Pop III stars discussed above, we verified that the estimated \textit{Strömgren} radii ($R_{\mathrm{S}}$) of individual star cells are consistently smaller than the adopted nebular radius in \textsc{Cloudy} ($R_{\mathrm{Cloudy}}$), which represents the physical extent of the local star-forming gas cloud. This ensures that the modeled Pop~III H\,\textsc{ii} regions remain compact and ionization-bounded. Moreover, the typical TNG50 gas cell size as shown in Table~\ref{tab:galaxy_data} ($\sim80$--$180\,\mathrm{pc}$) is systematically larger than the derived $R_{\mathrm{S}}$ values, indicating that such compact ionized regions are well contained within individual gas cells of the TNG50 galaxies. Therefore, our imposed size limit is both physically motivated and numerically consistent with the effective spatial resolution of the simulation.

\section{Spectral Analysis}
\label{spectral}
In this section, we present a detailed analysis of the integrated spectra obtained from the photoionization code \textsc{Cloudy}, focusing on the spectral characteristics of galaxies across different redshifts hosting Pop III stars. Figure~\ref{fig:intspe} presents the integrated spectra of the simulated galaxies in our sample, highlighting the main emission features that trace their stellar populations. The integrated spectrum of the galaxy SG2, shown in the top right panel of Fig.~\ref{fig:intspe}, serves as the initial reference case. This system is characterized by a redshift of \( z = 6.0 \) and a stellar population entirely composed of Pop III stars, as indicated by the ratio \( M_{*,\text{PopIII}}/M_{*,\text{total}} = 1.0 \). The spectrum shows several key features that are diagnostic of metal-free, primordial star formation. Some important emission lines are shown in red-shaded regions. 
 \begin{figure}
    \centering
    \includegraphics[width=0.4\textwidth]{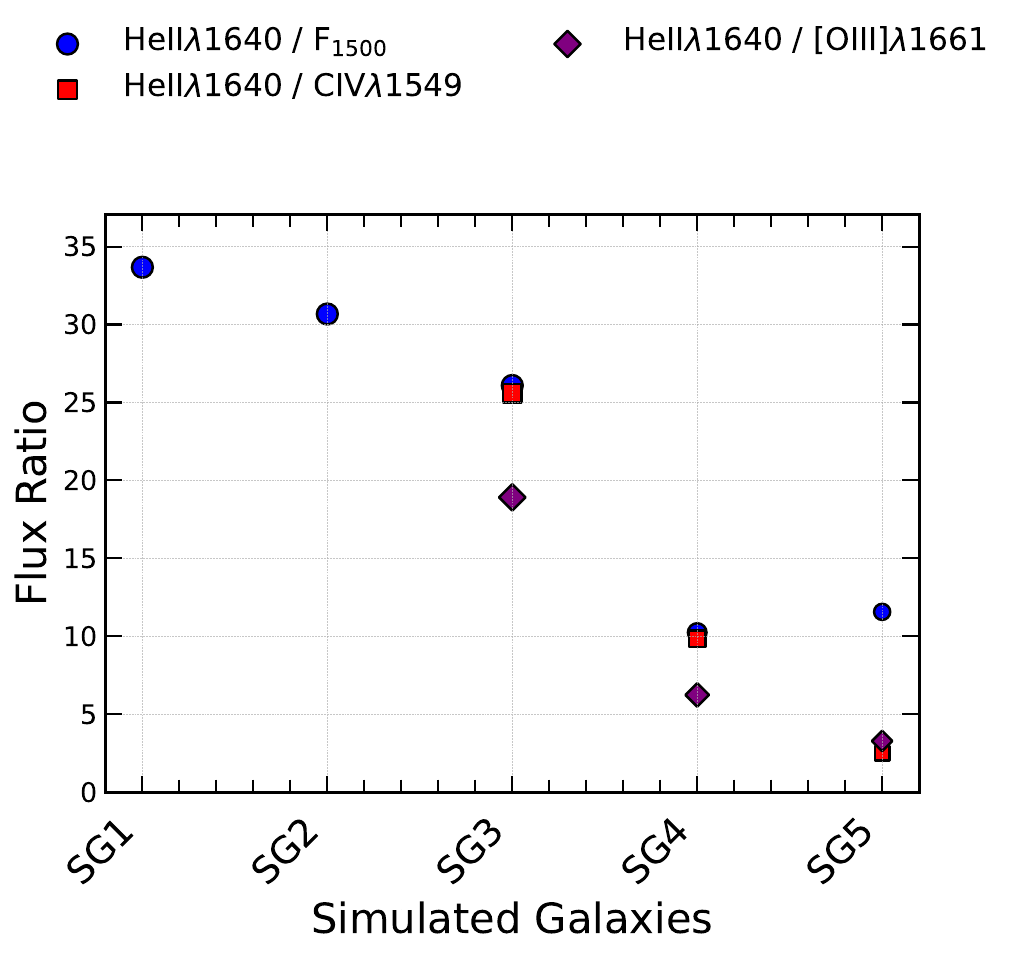}
    \caption{Flux ratios of the He\,\textsc{ii}$\lambda$1640 line with respect to Flux$_{1500}$, [OIII]$\lambda$1661 and CIV$\lambda$1549 for the five galaxies studied in this work. Galaxies SG2, SG3 and SG4 are at $z$=6 while SG1 is at $z$=10 and SG5 is at $z$=0. Since 100 \% Pop III systems like SG1 and SG2 don't produce metal lines such as [OIII]$\lambda$1661 and CIV$\lambda$1549, ratios
involving these lines are not shown for these galaxies. The ratio of $M_{*,\text{PopIII}}/M_{*,\text{total}}$, which decreases from SG1 to SG5, is represented by the downsizing of marker sizes.  }
    \label{fig:fluxratios}
\end{figure}
The most prominent feature in the UV range is the He\,\textsc{ii}$\lambda$1640 emission line. This emission line is considered to be a critical diagnostic tool  for identifying hot, metal-free stellar populations, as it originates from the recombination of highly ionized helium (He\(^{++}\)). The presence of this line, as explained earlier, is believed to be  a consequence of the hard ionizing spectra produced by Pop III stars, which are characterized by high effective temperatures and a lack of metals \citep{Tumlinson2000,Schaerer2002}. These stars are predicted to form from pristine, primordial gas composed of hydrogen and helium, as explained in Sect.~\ref{cloudy}, with no significant contribution from elements heavier than helium \citep{Bromm2002,Raiter2010}. The complete absence of metal lines such as [OIII]$\lambda$5007, [S II]$\lambda$6731, C III]$\lambda$1909 etc. in the spectrum of galaxy SG2  further validates its classification as a Pop III-dominated system, confirming the lack of significant metal enrichment in its stellar population. The galaxy SG1, at $z$=10 and with $ M_{PopIII}$/$M_{total}$ =1, has
a very similar spectrum to galaxy SG2, characterized by the absence of metal lines and slightly stronger hydrogen and helium emission features. SG2, however, since it lies at a lower redshift (
z=6), placing it at a later evolutionary stage with comparatively weaker H and He lines.
% In contrast, star-forming regions with even modest metal enrichment typically exhibit strong forbidden oxygen lines, such as [O III] 4959, 5007~\AA{}, due to efficient cooling processes in metal-rich gas.
Next, we examine the spectrum of galaxy SG3, shown in the middle left panel of Fig.~\ref{fig:intspe}, which exhibits a more complex stellar population. This system has a ratio \( M_{*,\text{PopIII}}/M_{*,\text{total}} = 0.8 \), indicating that 80\% of its stellar mass is composed of Pop III stars, while the remaining 20\% is contributed by Pop II stars. The presence of Pop II stars, which form from gas that has undergone some degree of metal enrichment, introduces detectable changes in the spectrum. Specifically, metal lines such as [S II]$\lambda$6731, C III]$\lambda$1909 and  the [O III] doublet at 4959 and 5007~\AA{} begin to appear. These lines are indicative of the presence of metals, which facilitate cooling, favouring the formation of lower-mass stars compared to Pop III stars \citep{Inoue2011,Nakajima2014}. The emergence of these features marks a transitional phase in the chemical evolution of the galaxy, where the first traces of metal enrichment are beginning to influence its spectral properties.\\
This trend becomes even more pronounced in the spectrum of galaxy SG4, shown in the middle right panel of Fig.~\ref{fig:intspe}. Here, the contribution from Pop II stars increases to 50\% of the total stellar mass (\( M_{*,\text{PopIII}}/M_{*,\text{total}} = 0.5 \)), resulting in a significant enhancement of the metal lines. The [S II]$\lambda$6731 and [O III]$\lambda$4959, 5007 lines are now much stronger, reflecting the increased metal content and the corresponding cooling processes in the interstellar medium. The presence of these lines underlines the growing influence of metal enrichment on the galaxy's spectral properties as the system transitions from a Pop III-dominated regime to one where Pop II stars play a more substantial role. Lastly, we show a galaxy, SG5, at $z$=0 which is
used as a proxy for local analogues of early galaxies that might be
hosting Pop III-like stars. We can notice that in addition to [S II]$\lambda$6731, C III]$\lambda$1909 and  the [O III] doublet at 4959 and 5007~\AA{} lines, more metal lines are visible in the integrated spectrum of this galaxy.  The higher metal enrichment, as expected in the local universe, can be seen through stronger metal lines in the emission line spectra. 

\section{Results}
\label{results}
In this section, we discuss some important emission line diagnostics that have been proposed to identify Pop III candidates \citep[e.g.][]{Feltre2016,Nakajima2022,Chrisholm2024}. 
The flux ratios of different spectral lines are plotted for each simulated galaxy in Fig.~ \ref{fig:fluxratios}. All models in this figure assume a Pop~III.1 IMF and spherical geometry and the impact of IMF and geometry are discussed later in this section.  As shown in Table~\ref{tab:galaxy_data}, galaxies composed exclusively of Pop III stars are represented by SG1 and SG2, whereas SG3, SG4, and SG5 show a gradual increase in the contributions of Pop II stars. The flux ratios computed here are  of He\,\textsc{ii}$\lambda$1640 / Flux$_{1500}$, He\,\textsc{ii}$\lambda$1640/CIV$\lambda$1549, and He\,\textsc{ii}$\lambda$1640/ [OIII]$\lambda$1661 which are useful diagnostic tools to differentiate between primitive Pop
III systems and more chemically evolved systems \citep{Schaerer2003,Nakajima2022}.

% Understanding the ionization states and chemical abundances in galaxies with different stellar populations depends on these ratios.
\begin{figure}
    \centering
    \includegraphics[width=0.5\textwidth]{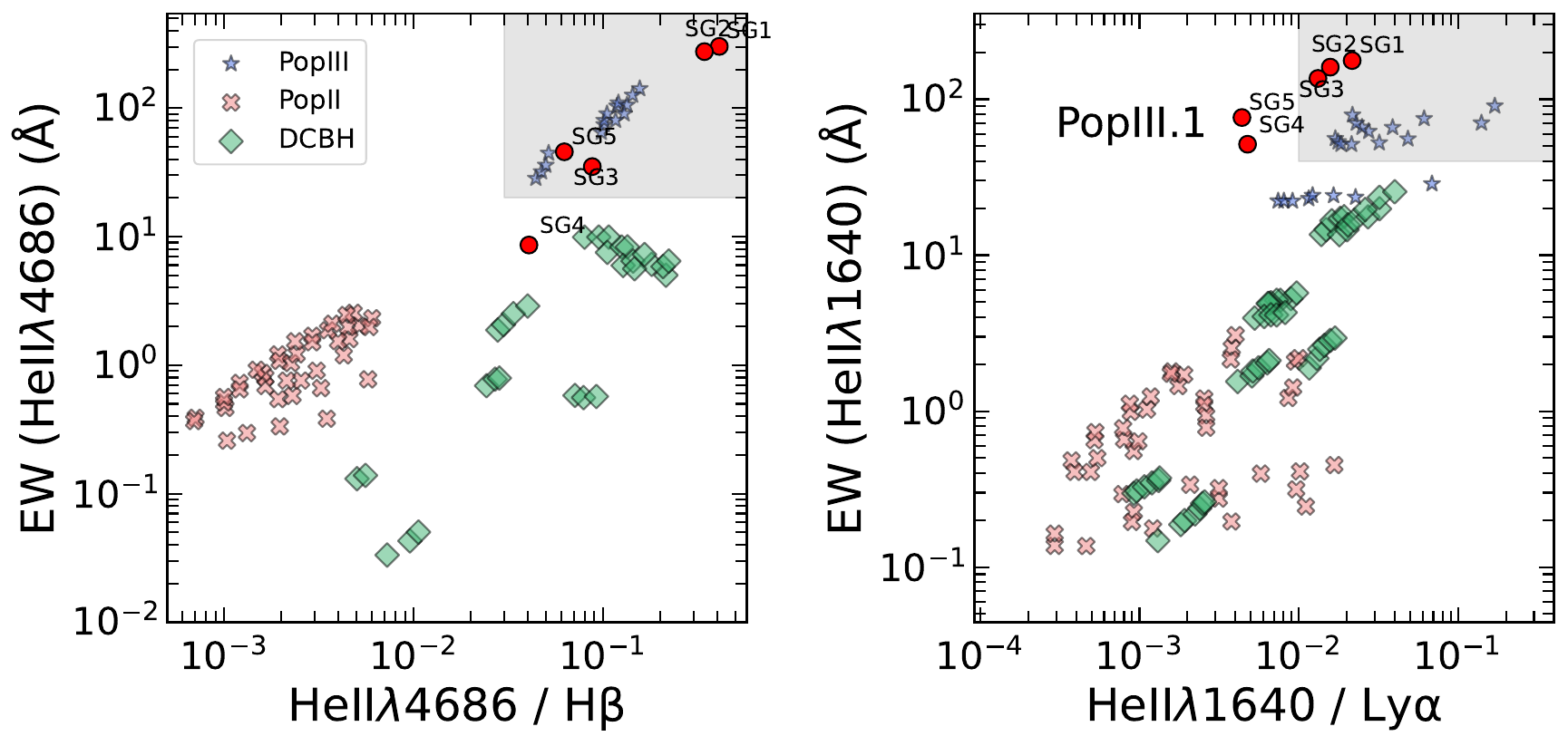}

\caption{
\textit{Left:} EW(He\,\textsc{ii}~$\lambda4686$) vs. He\,\textsc{ii}/H$\beta$ for different systems. The data shown as blue stars (Pop III), red crosses (Pop II), and green diamonds (direct collapse black holes, DCBHs) are taken from \citet{Nakajima2022}. The five simulated galaxies used in this work are shown as red circles. 
\textit{Right:} Same as the left panel, but for EW(He\,\textsc{ii}~$\lambda1640$) vs. He\,\textsc{ii}/Ly$\alpha$.
}

    \label{fig:Nakajima}
\end{figure}

\begin{figure}
    \centering
    \includegraphics[width=0.5\textwidth]{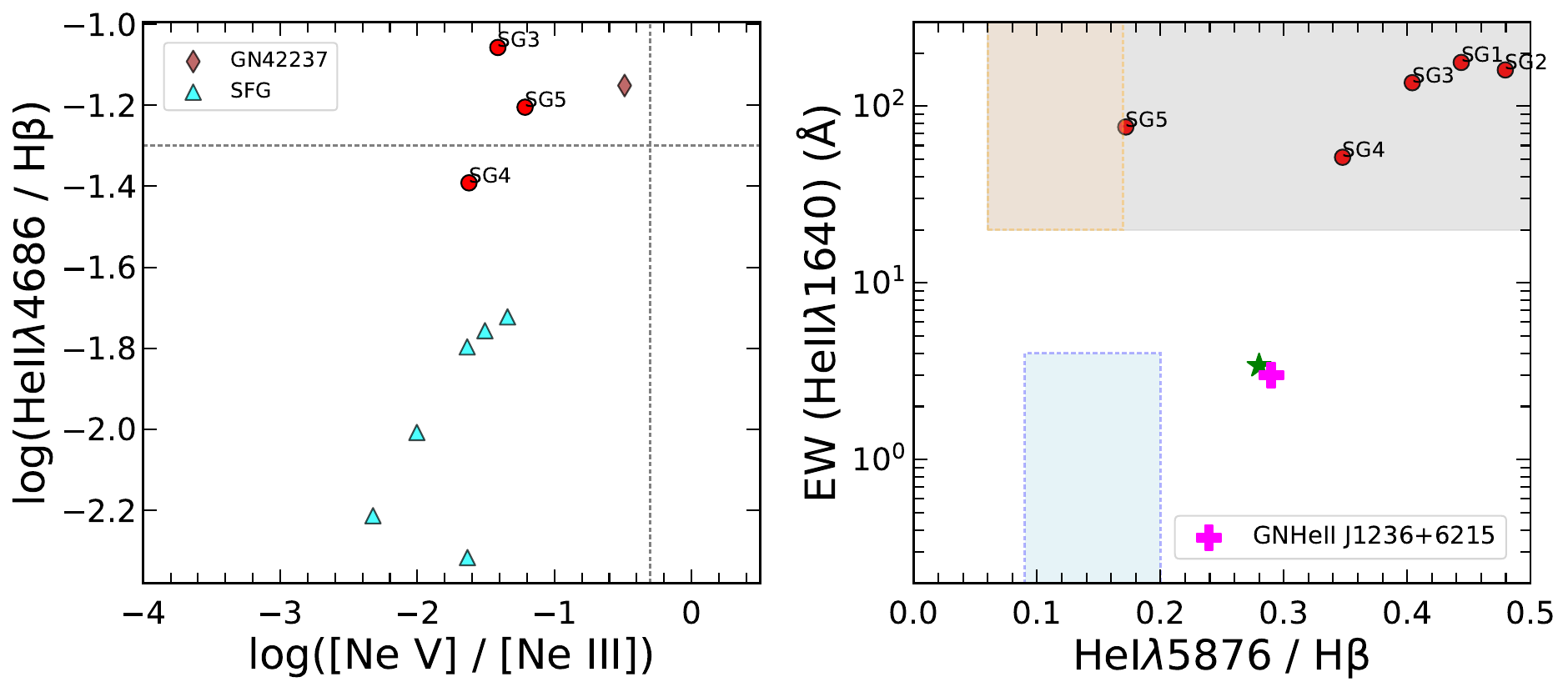}
    \caption{Left: The He\,\textsc{ii}$\lambda$4686/H$\beta$ vs. [Ne V]$\lambda$3427/[Ne III]$\lambda$3869 emission line diagnostic diagram. The grey horizontal line shown is proposed by \cite{Shirazi2012} and the vertical line instead is from \cite{Cleri2023} to  separate AGN and SF galaxies. The cyan triangles are a local sample of SF galaxies taken from \cite{Izotov2021}. GN 42 437 is denoted by the brown diamond and taken from \cite{Chrisholm2024}. Right: EW(He\,\textsc{ii}$\lambda$1640) vs. He\,\textsc{i}$\lambda$5876/H$\beta$ relation for the simulated galaxies. The orange and blue shaded regions denote the Pop III and Pop II model predictions from \cite{Nakajima2022}. The grey shaded region (which includes the orange region) is where galaxies hosting Pop III stars are predicted to lie, based on our models. The green star is a HeII emitter galaxy (object ID 50) taken from \cite{Raul2025}. The object GNHeII J1236+6215 is from \cite{Mondal2025}. }
    \label{fig:chrisholm}
\end{figure}
Fig.~\ref{fig:fluxratios} demonstrates notable differences in flux ratios based on the Pop III/Pop II mass fraction. Concerning the He\,\textsc{ii}$\lambda$1640 / Flux$_{1500}$ ratio, the highest
values correspond to the galaxies SG1 and SG2, which have $M_{*,\text{PopIII}}/M_{*,\text{total}}$
=1, while this ratio tends to decrease with decreasing the Pop III mass
fraction. This trend is expected since Pop III stars are predicted to
produce very strong H and He lines as a consequence of their extremely 
high effective temperature and absence of metals. The He\,\textsc{ii}$\lambda$1640 / Flux$_{1500}$ slightly drops for SG3 (20\% Pop II contribution), primarily due to a reduction in He\,\textsc{ii}$\lambda$1640 emission as the mass fraction of Pop III stars declines, which causes the flux ratio to decrease. Due to the contribution from Pop II star gas cells in this case, metal lines are visible, and so the He\,\textsc{ii}$\lambda$1640/CIV$\lambda$1549, and He\,\textsc{ii}$\lambda$1640/ [OIII]$\lambda$1661 flux ratios are also computed.  These ratios exhibit a more noticeable shift for SG4, which has a $M_{PopIII}$/$M_{total}$ 
=0.5, implying an increased contribution from Pop II stars in the mass fraction.   With the increase in the mass fraction of Pop II star cells (gas included), the production of metals also increases, causing the emission line intensities to be stronger, which causes the flux ratios to further decrease.  Lastly, SG5 is a galaxy that is more massive than all the others but has the lowest Pop III mass fraction (10\%). Because of this, the He\,\textsc{ii}$\lambda$1640 line produced has a relatively lower intensity, and the  He\,\textsc{ii}$\lambda$1640 / Flux$_{1500}$ ratio is partially higher than in SG4. Nevertheless, for SG5, the metal line intensities  are higher since Pop II star cells contain 90\% of their mass. Therefore, the flux ratios for He\,\textsc{ii}$\lambda$1640/CIV$\lambda$1549 and He\,\textsc{ii}$\lambda$1640/ [OIII]$\lambda$1661 are the lowest out of the five galaxies.
\begin{figure}
    \centering
    \includegraphics[width=0.48\linewidth]{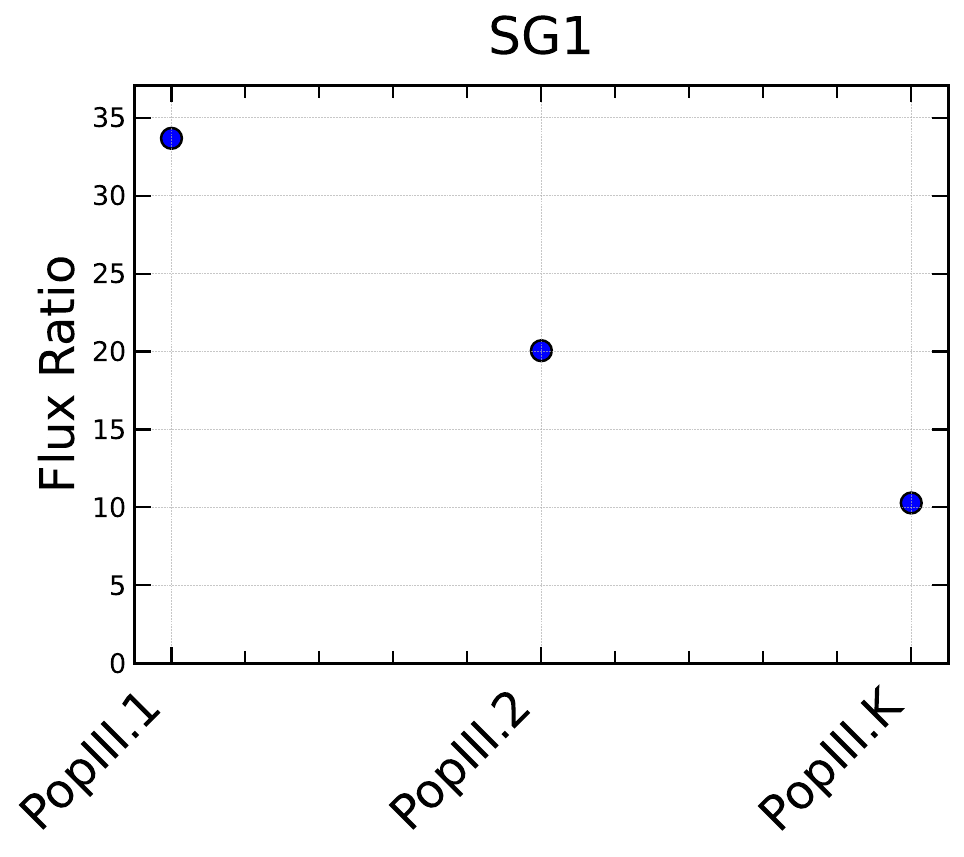}
    \includegraphics[width=0.48\linewidth]{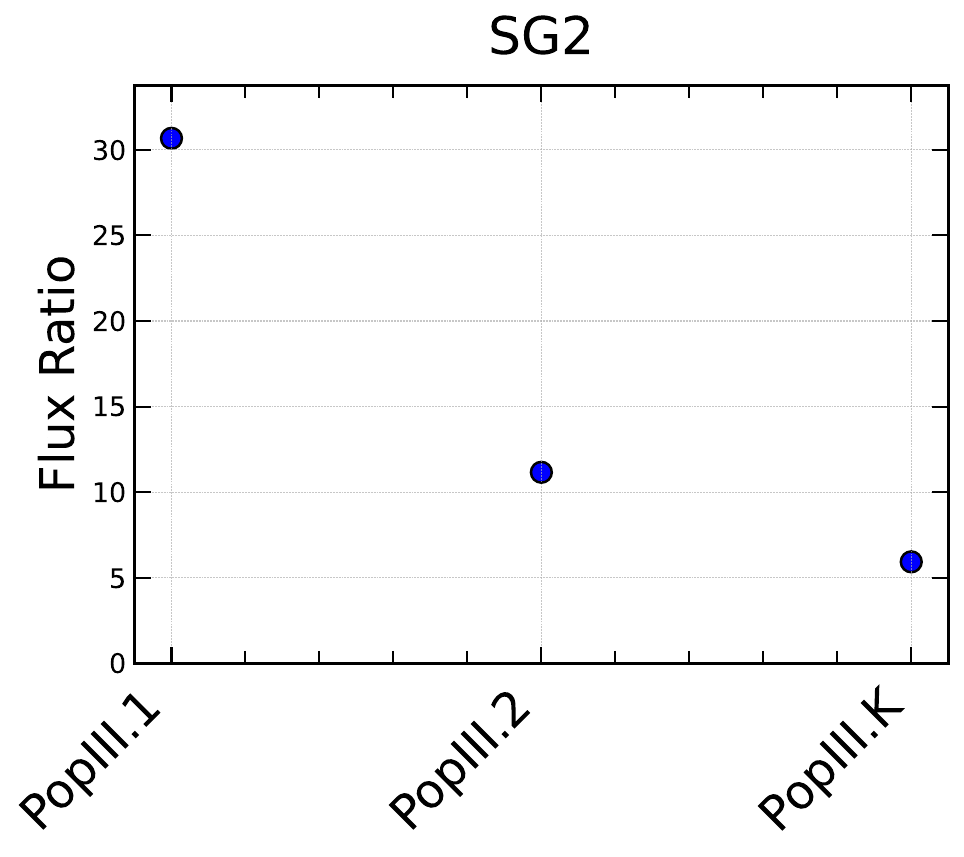}

    \includegraphics[width=0.48\linewidth]{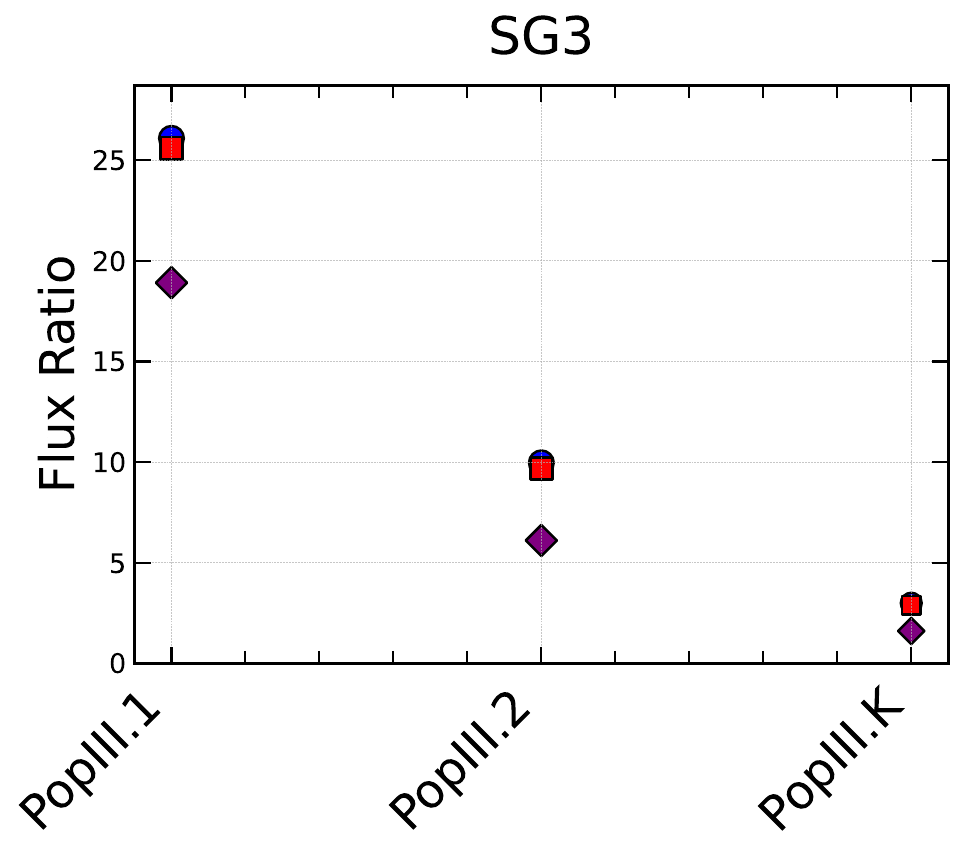}
    \includegraphics[width=0.48\linewidth]{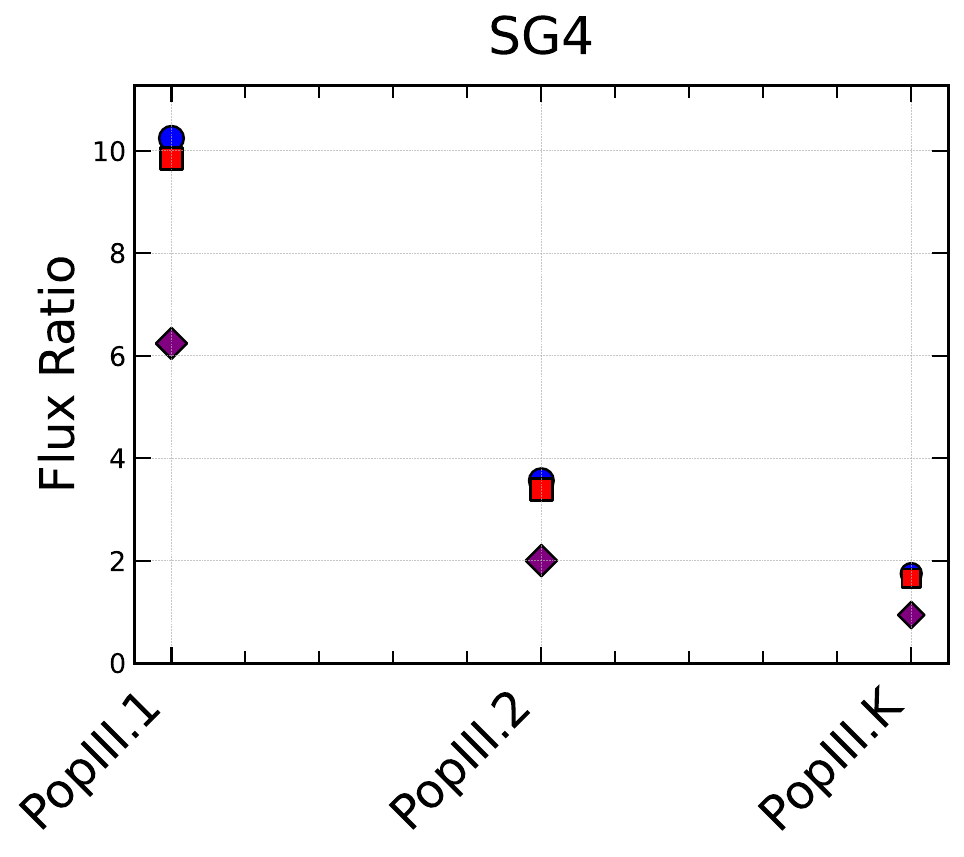}

    \caption{Effect of different IMFs on the flux ratios for our simulated galaxies. Each panel corresponds to a different galaxy; symbols are the same as in Fig.~\ref{fig:fluxratios}. The three different Pop III IMFs used in this work are shown on the x-axis. SG5 exhibits the same behavior and is therefore not shown for clarity }
    \label{IMFeffects}
\end{figure}

In Fig.~\ref{fig:Nakajima}, we show the relation between the EW(He\,\textsc{ii}$\lambda$4686) vs. the HeII/H$\beta$ and EW(He\,\textsc{ii}$\lambda$1640) vs. the HeII/Ly$\alpha$ which have been proposed by \cite{Nakajima2022} as a crucial diagnostic tool to identify Pop III systems. The gray shaded region in the top right part of both plots shows, according to this work, the area where Pop III galaxies are expected to exist, whereas Pop II and direct collapse black holes (DCBHs) as shown, populate other regions of the figure. It is evident that three of the five simulated galaxies used in this study—SG1, SG2, and SG3—with the highest Pop III contributions meet this requirement and are located in the shaded areas.  However, for the case of galaxies SG4 and SG5, which have 50\%  and 10\% contribution from  Pop III  stars, respectively, EW obtained for He\,\textsc{ii}$\lambda$4686 and especially He\,\textsc{ii}$\lambda$1640 fall outside the suggested shaded region of the plot. Hence, galaxies in which Pop III stars exist but may not be majorly dominated by Pop III stars might go undetected according to this criterion. To check the robustness of our results, we also rerun our analysis by restricting the final spectral reconstruction to stellar particles only younger than 10 Myr and and by assigning Pop II stellar particles to the nearest available Yggdrasil metallicity grid as mentioned in Sect.~\ref{cloudy}. The differences between the He\,\textsc{ii} EWs and HeII/ H$\beta$ line ratios obtained in the case of galaxies with the age cutoff do not exceed $\sim$0.04--0.06 dex, but greater differences (up to  $\sim$0.3- 0.9 dex) happen in those cases where old Pop II systems make a major contribution to the ionization spectrum. Importantly, even in these tests, the qualitative behaviour of the models and the relative location of our sources in the diagnostic diagrams remain unchanged .The same galaxies were also examined using the \cite{Chrisholm2024} log (HeII/H$\beta$) vs. log([Ne\textsc{v}]/[Ne\textsc{iii}])  diagnostic diagram, as shown in Fig.~\ref{fig:chrisholm}. All the galaxies in our sample that show these Ne lines in their spectra have higher HeII$\lambda$4686 / H$\beta$ and lie around the horizontal line prescribed by \cite{Shirazi2012} to distinguish between SF galaxies and AGN. Instead, a sample of local low-metallicity SF galaxies shown in cyan triangles occupies a different region \citep{Izotov2021}. Furthermore, we show the EW(He\,\textsc{ii}$\lambda$1640) vs. the HeI$\lambda$5876/H$\beta$ relation in the left panel of Fig.~\ref{fig:chrisholm}. All of our simulated galaxies are located in the upper grey shaded region, indicating that this diagnostic could be used as an additional tool to detect hybrid populations. We also tested this diagnostic with our models of plane-parallel geometry, and all the galaxies lie in the shaded region as well (see Sect.~\ref{Geometry}).   
 Both the  log([Ne\textsc{v}]/[Ne\textsc{iii}]) and the  HeI$\lambda$5876/H$\beta$  ratio may be valuable diagnostics for identifying systems hosting Pop III stars, especially in the context of hybrid stellar populations, as all of the galaxies in our sample that contain Pop III stars consistently occupy a similar region of the diagram, distinct from local SF galaxies.
 We note that while our models of the simulated galaxies do not take into account other mechanisms such as high-mass X-ray binaries \citep{Saxena2020}, faint or obsucred AGNs \citep{Feltre2016} etc., that can produce key diagnostic lines such as He\,\textsc{ii}$\lambda$1640 used here, future work incorporating these processes should be important to identify the origin of these high ionization features.  We discuss the possible caveats of this work later in this section. In any case, more research is required to develop improvements in other diagnostics that will enable the search for Pop III candidates even in complex and hybrid stellar population systems, which cosmological simulations like \textsc{IllustrisTNG} indicate are feasible across different redshifts.

\subsection{Impact of IMF on the flux ratios}
In this subsection, we discuss the possible effects of using different Pop~III IMFs on the emission line fluxes. Our post-processing framework allows us to investigate the effect of the Pop~III IMFs on nebular emission in a controlled parameter space. In  Fig.~\ref{IMFeffects} we show the flux ratios as a function of the three Pop~IIII IMFs used in this work and described in Sect.~\ref{method}.

The predicted flux ratios (symbols in Fig.~\ref{fig:fluxratios}) exhibit a clear dependence on the assumed Pop~III IMF. More top-heavy IMFs systematically produce higher He II–based flux ratios, with Pop III.1 being the most top-heavy yielding the largest values, followed by Pop III.2, and Pop III.K having a Kroupa IMF, producing the lowest ratios. This decline reflects the decreasing contribution of very massive stars and the associated softening of the ionizing spectrum in the three IMFs.

This IMF dependence directly impacts detectability. Based on the \cite{Nakajima2022} diagnostics, Pop III.1-dominated systems are the most readily detectable through their enhanced He II emission, as shown in Fig.~\ref{fig:Nakajima}.

\begin{figure}
    \centering
    \includegraphics[width=0.5\textwidth]{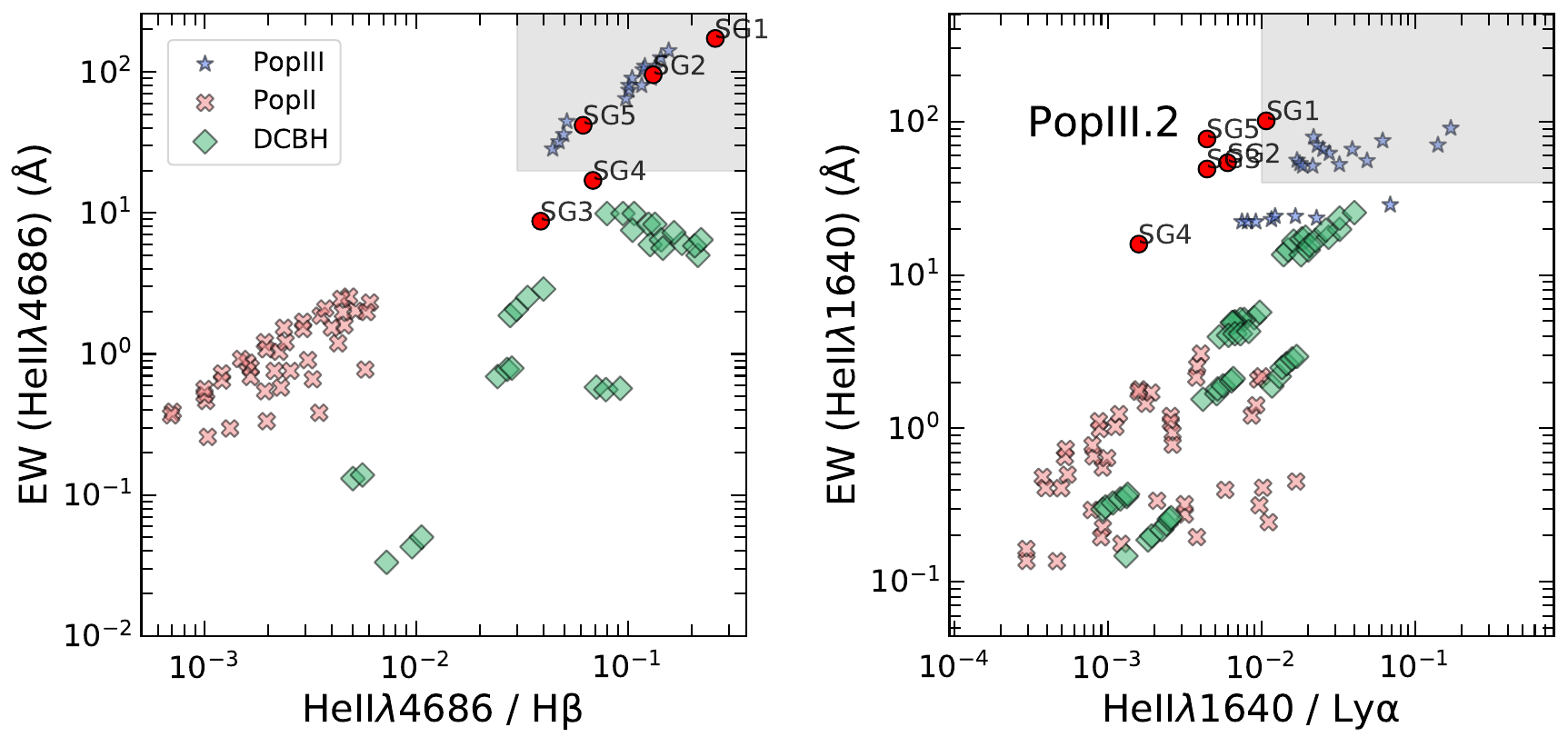}
    \includegraphics[width=0.5\textwidth]{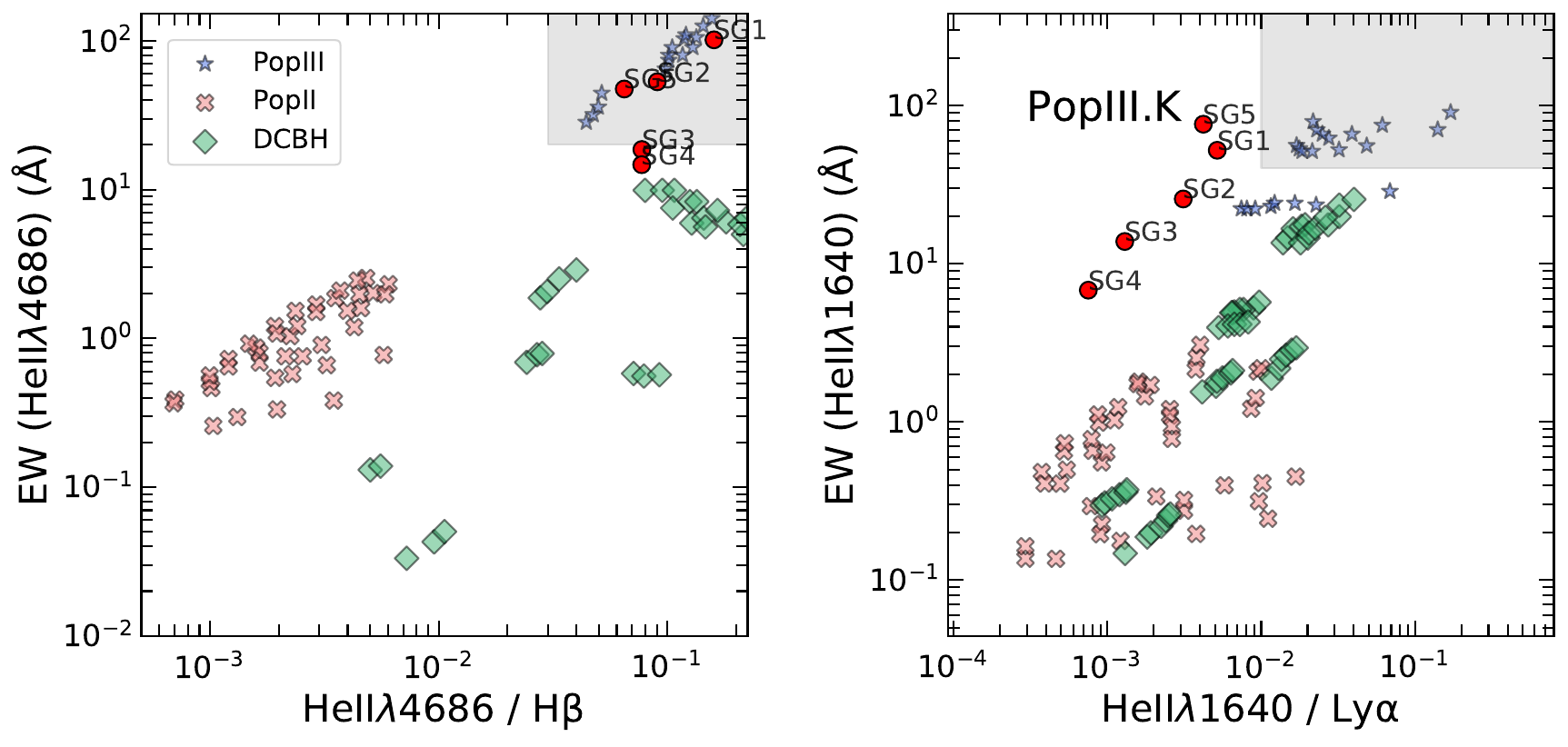}

\caption{
\textit{Left:} EW(He\,\textsc{ii}~$\lambda4686$) vs. He\,\textsc{ii}/H$\beta$ for different systems. \textit{Right:} Same as the left panel, but for EW(He\,\textsc{ii}~$\lambda1640$) vs. He\,\textsc{ii}/Ly$\alpha$. The top panels are related to models using the Pop~III.2 IMF whereas the bottom panels are for the case of Pop~III.K IMF.   The symbols shown are the same as in Fig.~\ref{fig:Nakajima}.}
    \label{fig:EW_Sphere}
\end{figure}

In Fig.~\ref{fig:EW_Sphere} we show the same diagnostic diagram for the case of Pop~III.2 and Pop~III.K IMFs. As shown in the top panel for the case of Pop~III.2 IMF, although most of the galaxies fall inside the shaded region based on the He\,\textsc{ii}$\lambda$4686 line, most of our simulated galaxies fall outside the criteria when the  He\,\textsc{ii}$\lambda$1640 line is employed. 

A similar trend is observed using the Pop~III.K IMF, where the He\,\textsc{ii}$\lambda$4686 line diagnostic performs better, while all of the galaxies lie outside the selection region when using He\,\textsc{ii}$\lambda$1640. This trend is expected given the less top-heavy nature of the Pop~III.K IMF, which produces a weaker extreme-UV ionizing spectrum resulting in lower He\,\textsc{ii}$\lambda$1640 emission.

\subsection{Impact of Geometry}
\label{Geometry}
\begin{figure}
    \centering
    \includegraphics[width=0.4\textwidth]{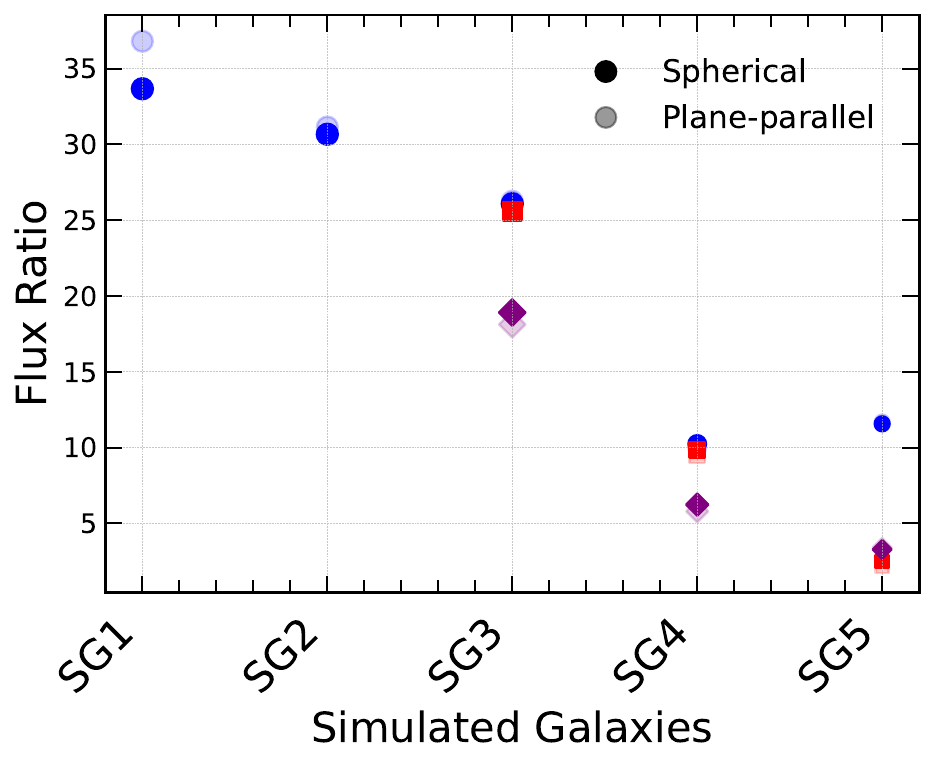}
    \caption{Same as Fig.~\ref{fig:fluxratios} but showing a comparison of flux ratios  between spherical and plane-parallel geometries.}
    \label{fig:fluxratiosgeo}
\end{figure}

In this subsection, we examine the impact of the assumed gas geometry in the \textsc{Cloudy} simulations by comparing results obtained using spherical and plane-parallel geometries. As shown in Fig.~\ref{fig:fluxratiosgeo}, the choice of geometry leads to systematic but modest differences in the predicted emission-line flux ratios for our simulated galaxies. However, the overall trends and qualitative behavior remain unchanged between the two geometries. 
For galaxies SG1 and SG2 that are totally dominated by Pop~III stars, the plane-parallel case shows slightly higher He\,\textsc{ii}$\lambda$1640 fluxes. As the fraction of Pop~III stars in the galaxies decreases the effect of geometry becomes relatively small.
These differences arise because of the ionization structure and treatment of the diffuse radiation field inherent to the two geometrical assumptions. In the plane-parallel case, the slab geometry maintains a relatively extended He$^{++}$ region, which can enhance UV recombination-line emission such as He\,\textsc{ii}$\lambda$1640. In contrast, spherical models account for the geometric dilution of the radiation field and this results in slightly more compact He$^{++}$ zone and modestly reduced He\,\textsc{ii}$\lambda$1640 emission. Such geometry-dependent effects are a well-known feature of photoionization modeling and are discussed in detail in the literature \citep{Schaerer1999,Ferland2013,Cloudy2017}.
This trend is also consistent in the \cite{Nakajima2022} diagnostic plots where the plane-parallel case is shown in Fig.~\ref{fig:Planepara}. The most top-heavy Pop~III IMF (PopIII.1) produces the highest He\,\textsc{ii}$\lambda$1640/Ly$\alpha$ ratios as shown in the figure. As the IMF becomes progressively less top-heavy going from top panel (Pop~III.1) to the bottom (Pop~III.K), the He\,\textsc{ii}$\lambda$1640/Ly$\alpha$ ratios decreases since the  production rate of He$^{+}$-ionizing photons decreases, leading to a contraction of the He$^{++}$ zone.

In contrast, He\,\textsc{ii}$\lambda$4686 is much less affected by the assumed geometry and so most of our galaxies lie in the grey shaded regions. Being an optical recombination line, its emissivity depends primarily on the total He$^{++}$ recombination rate and is less sensitive to the detailed spatial distribution of the ionized gas. As a result, variations in geometry lead to only minor changes in He\,\textsc{ii}$\lambda$4686, consistent with both our results and previous photoionization studies (e.g. \citealt{Cloudy2017}).

\subsection{Caveats and Limitations}
\label{caveats}
In this subsection, we outline the key assumptions in our modeling and discuss their possible impacts on the results of this work: \\

Firstly, we note that there is no explicit prescription for Pop~III star formation or its feedback in the \textsc{IllustrisTNG} simulations, including TNG50. Dedicated simulations of 
primordial star formation mentioned earlier, such as the Renaissance simulations \citep{Xu2016}, \texttt{FirstLight} \citep{Ceverino2017}, \texttt{THESAN} \citep{Kannan2022}, and the \texttt{dustyGadget}-based 
models of \citet{Venditti2023}, include different prescriptions of primordial chemistry, radiative transfer, metal enrichment, and Pop~III star formation prescriptions that are absent 
from TNG50. However, large-volume cosmological simulations that simultaneously provide publicly available galaxy catalogues and a fully self-consistent treatment of Pop~III star formation 
and feedback remain limited. We therefore adopt a post-processing approach that allows us to exploit the realistic galaxy environments and resolved local gas properties of 
TNG50 while explicitly exploring the impact of Pop~III stellar populations on observable emission-line diagnostics.Post-processing also provides the flexibility to test a wide range of SEDs for Pop III galaxy particles and local environments at the time of analysis without rerunning the full cosmological simulation, making it an efficient and physically consistent framework for such studies \citep[e.g.][]{Grisdale2021}. 

A major source of uncertainty is the IMF of Pop~III stars. Although TNG50 adopts a universal Chabrier IMF, theoretical models 
generally suggest that the IMF of Pop~III stars should be more top-heavy than more evolved stars \citep{Bromm2004,Hirano2014}. A more top-heavy IMF will lead to a harder ionizing radiation field, with
more He$^{+}$-ionizing photons produced, resulting in stronger nebular features like He\,\textsc{ii}\,$\lambda1640$. Hence, both the absolute line luminosities and the detectability of 
Pop~III galaxies could be different compared to what would be seen from a completely self-consistent hydrodynamical simulation. In this paper, we address this issue partly
by adopting the SEDs of Pop~III from the \textsc{Yggdrasil} model and varying the Pop~III mass fraction, while neglecting their dynamical impact.
Additionally, the transition from Pop~III to Pop~II star formation processes is sensitive to enrichment and mixing processes. Within TNG50, metals and feedback effects are modelled using 
subgrid physics recipes calibrated to galaxy evolution rather than primordial star formation \citep{Vogelsberger2013,Weinberger2017}. Specific studies aimed at Pop~III 
star formation processes indicate that the efficiency of metal mixing may play an important role not only in the duration of these processes but also in the total mass fraction 
$M_{\rm PopIII}/M_\star$ \citep{Sarmento2019,Venditti2023}. Higher enrichment levels inhibit future Pop~III activity, whereas low efficiency of metal-mixing allows pockets of pristine material to survive for an extended period. Thus, the adopted mass fractions of Pop~III stars should be considered as an exploratory parameter in our study.

 The presence of late Pop~III star formation episodes at $z \lesssim 10$ is currently highly uncertain and sensitive to the competition between enrichment,
feedback, gas infall, and mixing processes. Recent theoretical studies have suggested the continued occurrence of Pop~III star formation within chemically inhomogeneous regions until the end of the EoR \citep{Venditti2023,Venditti2026,Rusta2026}. However, different numerical models provide widely 
different predictions for the number of such systems. Since the TNG50 simulations do not model primordial chemistry and Pop~III feedback, we are not able to estimate the true number of such events accurately. We  therefore consider the extremely metal-poor stellar populations seen in TNG50 as Pop~III-like objects and study the 
consequences of their existence for observational studies. Our results thus have to be considered as a proof-of-concept study for mixed Pop~III/Pop~II diagnostics.

% Our modeling of the emission lines involves several assumptions that should be taken into account when interpreting the results. First, we adopt a top-heavy Pop~III IMF, which has also been used in many reionization studies, but not uniquely constrained; alternative IMFs (e.g. different slopes or upper-mass cutoffs) would change the ionizing photon budget and thus the predicted He\,\textsc{ii}/H$\beta$ ratios (see e.g. \citealt{Stanway2019}). 
% For e.g. \cite{Grisdale2021} showed in their work  when adopting different IMF slopes and upper-mass limits for Pop~III stars in post-processed simulations, the resulting He II emission can vary by more than an order of magnitude, with top-heavy IMFs producing more He II emission compared to more Salpeter-like slopes. Similarly, \citet{Raiter2010} and \citet{Schaerer2002} demonstrated that harder spectra from very massive stars ($>300,M_\odot$) yield higher He,\textsc{ii}/H$\beta$ ratios whereas shallower or truncated IMFs result in significantly weaker He,\textsc{ii} emission. Thus, while our adopted top-heavy IMF choice provides an upper limit on the ionizing output from Pop~III stars, if the IMF were less extreme, the He II emission would be lower than what is presented here.

Second, for Pop~II galaxy populations we assume a single metallicity although many observable stellar systems host a metallicity distribution and temporal evolution, and since ionizing photon hardness depends on metallicity, this assumption will change the resulting He II emission \citep{Gutkins2016,Steidel2016,Berg2021}. Our choice of  $Z=0.02 $ is a conservative upper bound designed to maximize the contribution of classical stellar channels—particularly WR stars to the production of HeII-ionizing photons \citep{Schaerer2003,Crowther2007}. At lower metallicity, the WR contribution declines sharply, and the He II/H$\beta$ ratio from Pop II populations would decrease \citep{Schaerer2003,Crowther2007,Eldridge2008}. Thus, our models provide an upper bound on the expected Pop II He II contribution—using lower-Z Pop II populations would only increase this discrepancy and will produce further lower He II/H$\beta$ ratios.

Finally, we do not include non-stellar high-energy ionizing sources such as X-ray binaries (XRBs) in our models. At very low metallicities, XRBs can contribute to additional hard photons \citep[e.g.][]{Fragos2013, Schaerer2019,Sartorio2023}, and recent work with the \textsc{XBPASS} models \citep{2025MNRAS.542.2087B} show that including XRBs increases the He$^{++}$-ionizing photon flux by factors of a few. However, even when XRBs are included, the resulting He\,\textsc{ii}/H$\beta$ ratios (see Fig. 18 of \citeauthor{2025MNRAS.542.2087B} \citeyear{2025MNRAS.542.2087B} ) remain well below those produced by Pop~III-dominated populations, typically reaching at most He\,\textsc{ii}/H$\beta \sim 10^{-3}$. Hence, while XRBs can (modestly) enhance the ionizing continuum at low metallicities, they cannot by themselves reproduce the high nebular He\,\textsc{ii} intensities characteristic of nearly metal-free stellar populations \citep{Senchyna2020}.

\subsection{Comparison with Recent Pop III Modelling and JWST Observations}

Recent theoretical and observational studies show growing evidence that Pop~III star formation continues even until the end of the reionization epoch in chemically inhomogeneous conditions.  \citet{Venditti2023} used   \texttt{dustyGadget}, a hydrodynamical code \citep{Graziani2020} including inhomogeneous metal enrichment and Pop~III star 
formation prescriptions to show that Pop~III star formation ($\sim10^{-3.4}-10^{-3.2}\,\mathrm{M_{\odot}\,yr^{-1}\,cMpc^{-3}}$) can continue at $z\sim6-8$ within pristine pockets embedded in globally enriched environments . Moreover, \citet{Venditti2023} showed that inefficient mixing of metals plays an important role 
in Pop~III star formation and shows a Pop III/ Pop II mass fraction of $\lesssim$ 0.1 per cent is found in 10 percent of the massive galaxies having masses  $M_{*} \lesssim 3 \times 10^9  M_\odot$.

More recent studies have extended this framework to investigate the nebular signatures and detectability of Pop~III-forming regions in hybrid Pop~III/Pop~II systems 
\citep{Venditti2026}.  They followed a similar approach combininig cosmological simulations with detailed \textsc{Cloudy} photoionization 
modelling to study compact Pop~III star-forming pockets embedded within enriched galaxies at $z\sim6.5-9$. Their fiducial models considering young Pop~III bursts with ages  $\lesssim1\,{\rm Myr}$, top-heavy IMFs, and Pop~III stellar masses of order $M_{\rm III}\sim10^{5}-10^{6}\,{\rm M_\odot}$ show the strongest He II 1640 emission. The nebular calculations adopted dense gas 
conditions with hydrogen number densities
\[
n_{\rm H}\sim10^{3}-10^{4}\,{\rm cm^{-3}},
\]
 motivated by compact high-redshift star-forming clumps.

Similarly,  recently identified ``Hebe'' source near GN-z11 by \cite {Maiolino2026} showed that using a combination of stellar population synthesis and nebular 
emission modelling the peculiar system can be explained. Their analysis favoured a scenario involving  Pop~III stars where they use models with dense nebular environments spanning a range of $n_{\rm H}\sim10^{3}-10^{6}\,{\rm cm^{-3}}$ using Pop III and hybrid models from \cite{Nakajima2022} and \cite{Rusta2026} respectively.

In this work we adopt a complementary approach in modeling, which is also a more exploratory one. Due to the relative unavailability of large volume cosmological simulations in which the formation and radiative feedback of Pop~III stars are considered self-consistently, we use  the \textsc{IllustrisTNG50} simulation and  employ postprocessing to implement 
Pop~III population using \textsc{Yggdrasil} SEDs and \textsc{Cloudy} photoionization analysis tools.  We again emphasize that our study should be taken as more of a proof-of-concept study to examine the impact of Pop~III and Pop~II stars on some diagnostic lines, rather than a fully predictive numerical model. 

Nevertheless, the physical properties investigated in our calculations generally match those derived in the recent studies involving Pop III galaxies mentioned above.
For instance, across all galaxies analyzed in this work, the inferred densities span approximately
$n_{\rm H}\sim10^{2.3}-10^{4.2}\,{\rm cm^{-3}}$, with most systems clustering around
$n_{\rm H}\sim10^{3.8}-10^{4.1}\,{\rm cm^{-3}}$.
These values are broadly consistent with the dense, compact nebular environments inferred in recent JWST studies of extreme emission-line galaxies and He\,\textsc{ii}-emitting 
candidates at high redshift \citep{Maiolino2026,Ubler2026}. 
\begin{figure}
    \centering
    \includegraphics[width=0.4\textwidth]{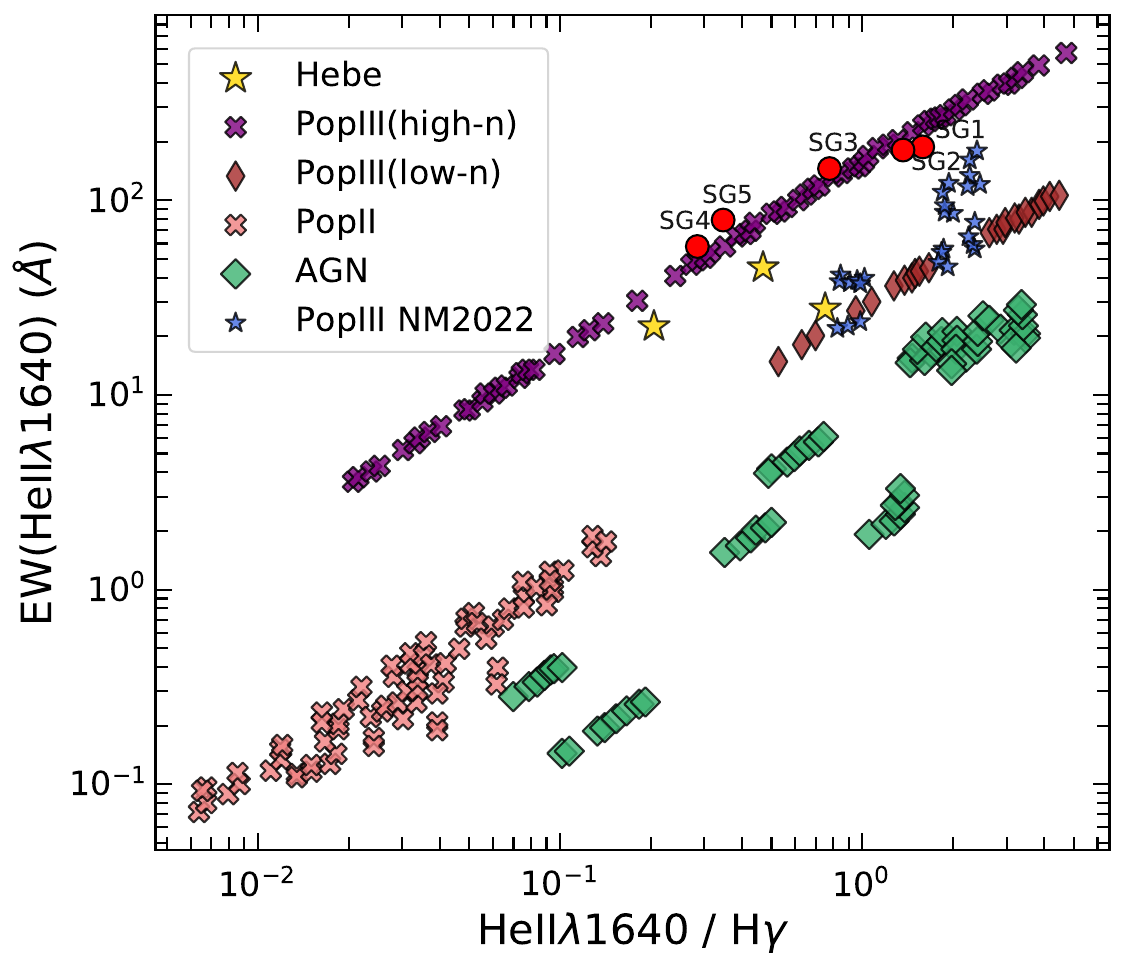}
    \caption{
EW of He\,\textsc{ii}\,$\lambda1640$ as a function of the He\,\textsc{ii}\,$\lambda1640$/H$\gamma$ line ratio for different ionizing sources.  The JWST Pop III candidate \textit{Hebe} (total) and its two components are shown in yellow stars, whereas photoionization models from \citet{Nakajima2022} are shown in blue stars (Pop III), red crosses (Pop II), and green diamonds (AGN). The five simulated galaxies used in this work are shown as red circles. Pop III and hybrid models from \citep{Rusta2025} are shown in purple crosses (Pop III high-n) and brown diamonds (Pop III low-n). }
    \label{fig:Hebe}
\end{figure}
We demonstrate this compatibility in Fig.~\ref{fig:Hebe} showing the EW of HeII$\lambda1640$ as a function of the HeII$\lambda1640$/H$\gamma$ line ratio . Here we show that our galaxies exhibit a comparable region in this diagram compared to both the Pop~III and hybrid Pop~III/Pop~II galaxy models discussed by \citet{Rusta2026} and \citet{Nakajima2022}. In particular, our mixed-population systems overlap with the high-ionization, weak-metal-line regime associated with  compact Pop~III-forming pockets and recent JWST Pop~III candidates, further supporting the interpretation that even small localized Pop~III components can significantly  affect nebular line diagnostics.
\section{Discussion and Conclusions}
\label{conclu}
In this work, we present a proof-of-concept study of Pop III galaxy emission line characteristics using simulated galaxies from the \textsc{IllustrisTNG50} simulations.  Cosmological simulations and other recent studies \citep[e.g.][]{Ciardi2005,Liu2020,Venditti2023,Venditti2026} show that complex systems of multiple stellar populations containing both Pop III and Pop II stars could co-exist over a range of redshifts. We use a set of  criteria to select galaxies from the \textsc{IllustrisTNG-50-1} in order to study how the emission line diagnostics of Pop III stars might change for such hybrid systems and challenge their detection .  First, we adopt the critical 
metallicity,  $Z_{crit}$ =0.02 $Z_{\odot}^{2}$  to distinguish between Pop III or Pop II star particles in galaxy simulations and select galaxies with different ratios $M_{*, PopIII}$ and $M_{*,total}$ which represent the total mass of Pop III stars and the galaxy's total stellar mass, respectively. We select five galaxies in which SG1 and SG2 totally consist of Pop III stars at $z$=10 and $z$=6 respectively, whereas SG3, SG4, and SG5 have a $M_{*,PopIII}$ /$M_{*,total}$ of 0.8, 0.5, and 0.1, respectively.  In this respect, our findings are aimed at providing qualitative indications about the sensitivity of diagnostics rather than quantitative predictions on the number of Pop~III star-forming galaxies .

Using distinct IMFs for Pop III and Pop II particles, we model every star particle in a galaxy using \textsc{Yggdrasil} models. Our post-processing method allows us to explore multiple modelling assumptions including variations in Pop~III IMFs and nebular geometries without the need to re run expensive cosmological  simulations. The emission line spectrum for each particle is obtained by running \textsc{Cloudy} photoionization models using the particle's total luminosity ($L_{\text{tot}}$) and the hydrogen number density ($n_{H}$) of its host cell. This enables a self-consistent modeling approach in which the local physical properties of each simulation cell are used to derive the corresponding emission line characteristics. These spectra are then integrated to obtain the stacked integrated spectra of a galaxy. In Fig.~\ref{fig:intspe}, we demonstrate how the features of the integrated spectrum vary based on the $M_{*,PopIII}$ /$M_{*,total}$. Galaxies dominated by Pop III stars show stronger H and He lines but negligible or no metal lines. Metal line intensities increase  as the $M_{*, PopIII}$ /$M_{*, total}$ decreases, and more Pop II stars contribute to heavier element enrichment.  In Fig.~\ref{fig:fluxratios}, we present different flux ratios where SG1 and SG2 show high He\,\textsc{ii}$\lambda$1640 / Flux$_{1500}$ ratios indicative of Pop III-dominated galaxies, whereas for SG3, SG4, and SG5, the flux ratios gradually decrease. As the ratio of Pop II stars increases in the system, more metal lines, such as [OIII]$\lambda$5007 etc, are visible, causing the flux ratios to decrease. We further demonstrate that the He\,\textsc{ii}$\lambda$1640 / Flux$_{1500}$  ratio increases with redshift, as shown by SG1 and SG2,  both galaxies with comparable stellar masses. Conversely, flux ratios with respect to metal lines decrease with a decrease in  redshift, as shown by SG4 and SG5, in accordance with progressive metal enrichment with time. We find that variations in the Pop~III IMF primarily affect the flux ratios as shown in Fig.~\ref{IMFeffects}, with the most top-heavy IMF producing the highest flux ratios based on He\,\textsc{ii}$\lambda$1640.    \\
% EW(He\,\textsc{ii}$\lambda$4686) vs. the HeII/H$\beta$ and EW(He\,\textsc{ii}$\lambda$1640) vs. the HeII/Ly$\alpha$ have been proposed by \cite{Nakajima2022} to be essential diagnostics to distinguish between Pop III and other systems.

In Fig.~\ref{fig:Nakajima} we show that when using the Pop~III.1 IMF, among the five galaxies selected here,  SG1, SG2, and SG3 fulfil the EW(He\,\textsc{ii}$\lambda$4686) vs. the HeII/H$\beta$ and  HeII/Ly$\alpha$ criteria suggested by \cite{Nakajima2022} for distinguishing Pop III galaxies and fall in the shaded region of the plot. SG4 and SG5, which have a $M_{*,PopIII}$ /$M_{*,total}$= 0.5 and 0.1, respectively, fall outside the expected region of Pop III galaxies. The variation of IMFs has less effect on the detectability based on the He\,\textsc{ii}$\lambda$4686 line, however as we move to less top-heavy IMFs as shown in Fig.~\ref{fig:EW_Sphere}, most galaxies fall outside the shaded region based on He\,\textsc{ii}$\lambda$1640 diagnostics. These UV lines such as He\,\textsc{ii} are also more sensitive to nebular geometry. Plane-parallel models generally predict higher He\,\textsc{ii}$\lambda$1640/Ly$\alpha$ ratios due to a more extended He$^{++}$ region compared to spherical geometries. This effect is shown in Fig.~\ref{fig:Planepara} where our galaxies are more shifted towards higher He\,\textsc{ii}$\lambda$1640/Ly$\alpha$ compared to the spherical case (Fig.~\ref{fig:Nakajima}). The optical He\,\textsc{ii}$\lambda$4686 line on the other hand is less effected by the geometry and most of our galaxies still fall in the shaded region. 

We notice that most galaxies, particularly those that are very young, where strong hydrogen and helium emission lines are present and metal lines are absent, consisting primarily  of Pop III stars, are expected to meet the distinct criteria proposed by \cite{Nakajima2022}. However, for galaxies in which the spectra show metal lines, we propose using the log([Ne\textsc{v}]/[Ne\textsc{iii}]) diagnostic as a complementary tool for identification. Furthermore,  all our galaxies lie in the upper end of the  EW(He\,\textsc{ii}$\lambda$1640) vs. the HeI$\lambda$5876/H$\beta$ diagram, providing an additional diagnostic pathway for identifying and characterizing galaxies with mixed stellar populations, which include both Pop III and Pop II stars. 
% In a forthcoming detailed work the role of Pop III IMF will be studied, exploring the dependence of the Pop III loci of these emission line diagnostic diagrams on the adopted IMF.\\
Owing to the flexibility of our post-processing approach, physically motivated spectral templates can be constructed spanning a wide range of stellar population mixtures using different IMFs and nebular conditions. This can be particularly helpful for interpreting the increasingly diverse and sometimes unexpected emission-line properties revealed by recent \textit{JWST} observations. While our models are not calibrated for reproducing any specific source, we show in Fig.~\ref{fig:Hebe} that our galaxies from our simulations fall into the same parameter regime occupied by recent Pop~III and Pop~III/Pop~II models in the literature , which have also been used to describe some Pop~III candidate systems, such as the HeII emitter Hebe near GN-z11 \citep{Nakajima2022,Rusta2026,Maiolino2026}.
In any case, in order to trace the earliest generations of stars more precisely, these diagnostics should be optimized further using both theoretical modeling and high-resolution observations coming from present and future facilities such as \textit{JWST} and ELT.

\section*{Acknowledgements}
SG, JVM, CK and JIP acknowledge financial support from the  State Agency for Research of the Spanish MCIU through the Center of Excellence Severo Ochoa’ award to the Instituto de Astrofísica de Andalucía CEX2021- 001131-S funded by MCIN/AEI/10.13039/501100011033, and from grant PID2022- 136598NB-C32 “Estallidos8. AF acknowledges support from the ERC Advanced Grant INTERSTELLAR
H2020/740120 (PI: Ferrara). Partial support (AF) from the Carl Friedrich von SiemensForschungspreis der Alexander von Humboldt-Stiftung Research Award is kindly acknowledged.

%%%%%%%%%%%%%%%%%%%%%%%%%%%%%%%%%%%%%%%%%%%%%%%%%%

\section*{Data Availability}

The simulations used in this work are based on the publicly available \textsc{IllustrisTNG} project (\url{https://www.tng-project.org/data/}). The photoionization models and post-processing scripts developed for this study will be shared upon reasonable request to the corresponding author.



\bibliographystyle{aa}
\bibliography{biblio} % if your bibtex file is called example.bib

\appendix
\section{\textsc{Yggdrasil}  and Photoionization models }
%\subsection{ \textsc{Yggdrasil} models}

Here we give further details about the stellar population synthesis and photoionization models used in the present study. In this regard, we demonstrate the evolution of the ionizing spectra used in this work to model the Pop~III stellar population and explore the dependence of our results on the gas geometry assumed.
\begin{figure}
    \centering
    \includegraphics[width=0.3\textwidth]{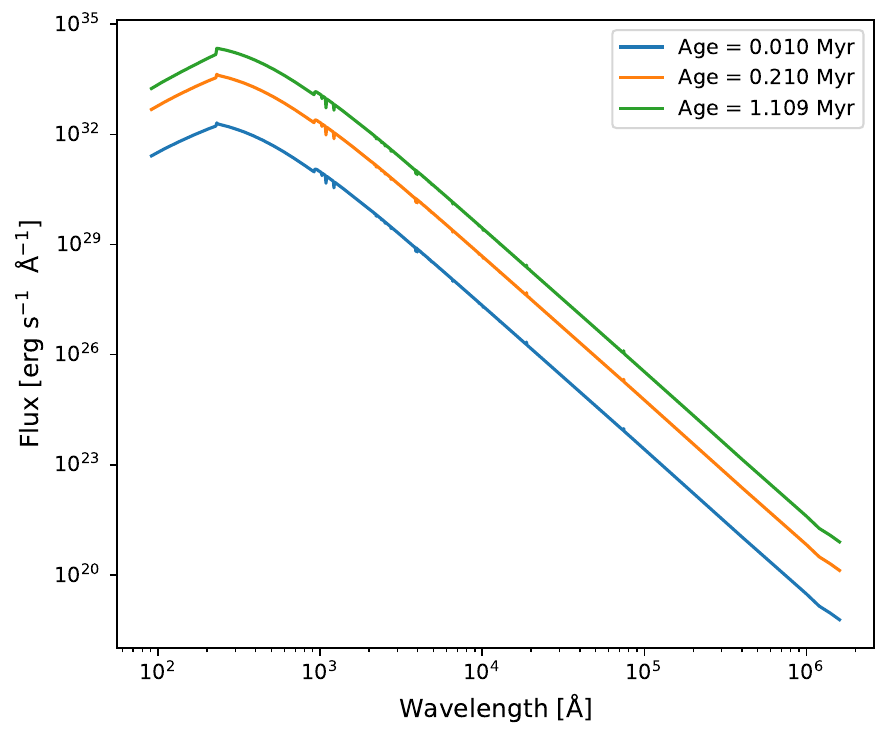}
    \caption{SEDs for the PopIII.1 for three different ages 0.010 Myr, 0.210 Myr, 1.109 Myr shown in
blue, orange, and green solid lines respectively.}
    \label{fig:grasil}
\end{figure}

Figure~\ref{fig:grasil} shows the PopIII.1 SEDs generated with \textsc{Yggdrasil} at three representative ages. The youngest population ($0.01\,{\rm Myr}$) exhibits an extremely hard ionising spectrum extending well beyond the He$^{+}$ ionization edge at 54.4 eV, reflecting the contribution of very massive metal-free stars. As the stellar population goes from younger to older, the ionizing photon production rate decreases rapidly due to the short lifetimes of the most massive stars, leading to a substantial softening of the spectrum. This evolution directly affects the strength of nebular recombination lines such as He\,\textsc{ii}\,$\lambda1640$ and He\,\textsc{ii}\,$\lambda4686$ as shown in Fig.~\ref{fig:EWAge}, which are highly sensitive to the availability of photons capable of doubly ionizing helium. Due to the top-heavy IMF, the SED for PopIII.1 becomes negligible for $t > 3.6 \times 10^6 $ Myr. All SEDs are for a
stellar population with a total mass of $10^5 M_{\odot}$.
\begin{figure}
    \centering
    \includegraphics[width=0.5\textwidth]{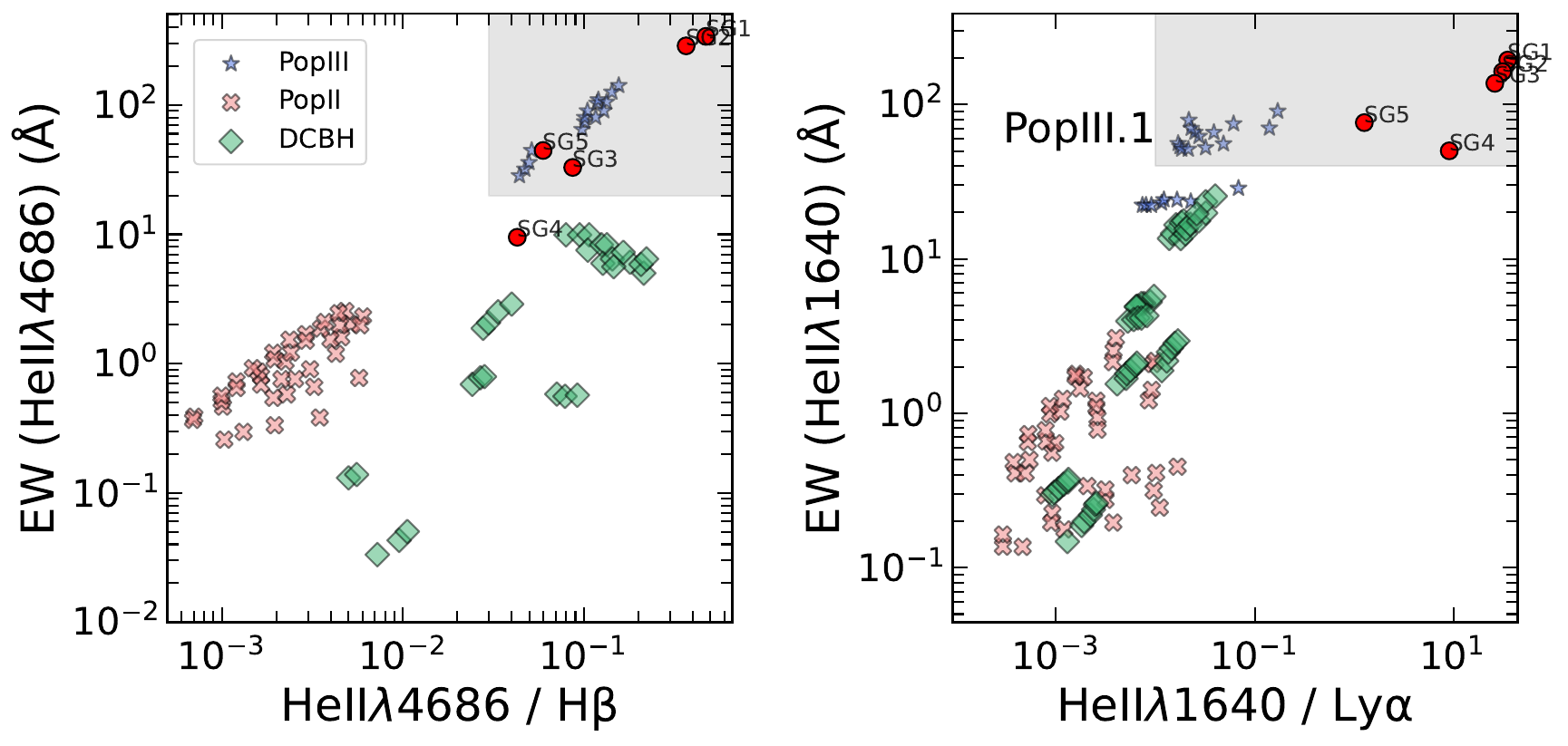}
    \includegraphics[width=0.5\textwidth]{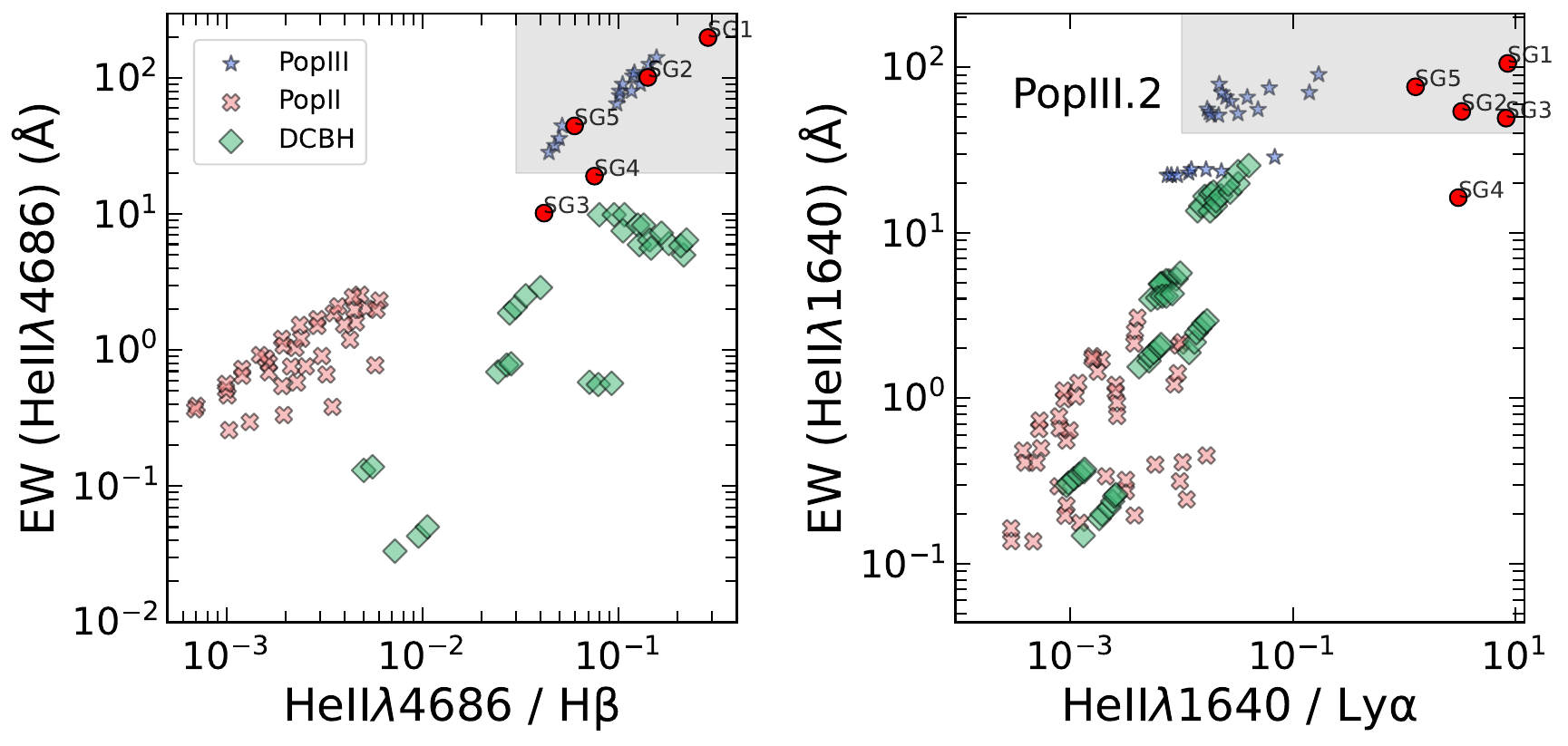}
    \includegraphics[width=0.5\textwidth]{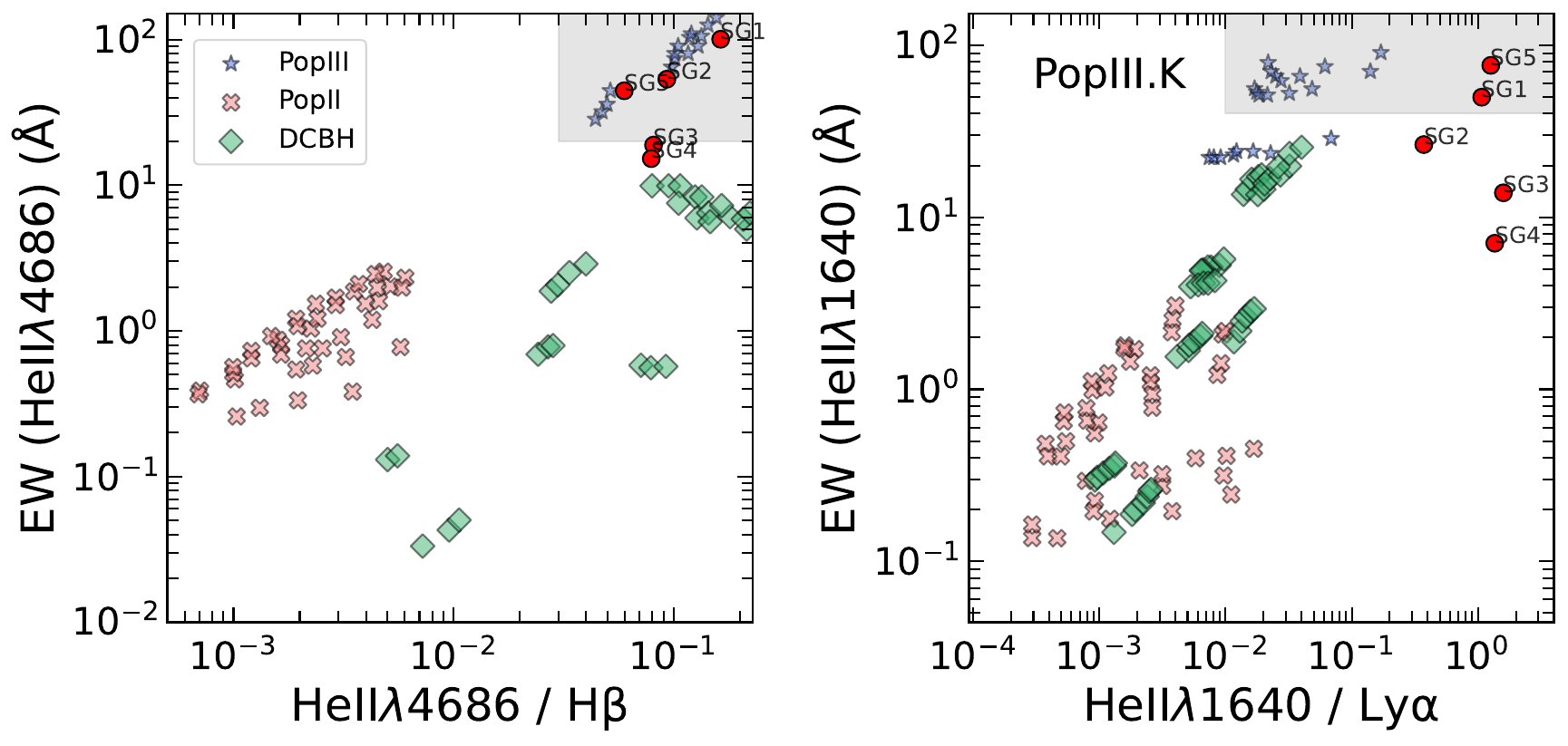}

\caption{
Same as Fig.~\ref{fig:EW_Sphere}, but for the case of models with plane-parallel geometry.
}

    \label{fig:Planepara}
\end{figure}

On the other hand, to evaluate the dependence of the predicted emission line diagnostics on the nebular structure and geometry, we additionally run the same simulated galaxies with plane-parallel photoionization models. Figure~\ref{fig:Planepara} presents the resulting EW versus emission line ratio diagrams for the three Pop~III IMF prescriptions considered in this work. While the detailed locations of individual models vary as a consequence of changes in the ionization structure and photon escape probabilities, the overall behaviour remains similar to that obtained for spherical geometries (Fig.~\ref{fig:Nakajima} and \ref{fig:EW_Sphere}). Because of a longer He$^{++}$ region than spherical geometries, plane-parallel models often predict larger He\,\textsc{ii}$\lambda$1640/Ly$\alpha$ ratios. Fig.~\ref{fig:Planepara} illustrates this effect, showing that our galaxies are more skewed towards higher He,\textsc{ii}$\lambda$1640/Ly$\alpha$ compared to the spherical case (Fig.~\ref{fig:Nakajima}). Overall, we show that the qualitative diagnostic trends identified in the main text are robust against reasonable variations in the assumed nebular geometry.
\label{lastpage}
\end{document}